\documentclass[twocolumn,twocolappendix]{aastex631}

\usepackage{amsmath,amstext}
\usepackage[utf8]{inputenc}
\usepackage[T1]{fontenc}
\usepackage{footmisc}
\usepackage{blindtext}
\usepackage{bm}

\newcommand{\D}{$^\circ$}
\newcommand{\per}{$^{-1}$}
\newcommand{\pers}{$^{-2}$}
\newcommand{\pert}{$^{-3}$}
\newcommand{\kms}{\mbox{km~s$^{-1}$}}

\newcommand{\msun}{M$_\odot$}
\newcommand{\lsun}{L$_\odot$}
\newcommand{\Ha}{H$\alpha$}
\newcommand{\HI} {\mbox{\rm H\textsc{i}}}
\newcommand{\HIsect}{H{\scriptsize I}}
\newcommand{\HII} {\mbox{\rm H\textsc{ii}}}

\newcommand{\htwo}{H$_2$}
\newcommand{\CII}{[C\textsc{ii}]}
\newcommand{\CIIsect}{[C{\scriptsize II}]}
\newcommand{\Cp}{C$^+$}
\newcommand{\NII}{[N\textsc{ii}]}
\newcommand{\OIII}{[O\textsc{iii}]}
\newcommand{\NCp}{$N_{\rm C^{+}}$}
\newcommand{\NCpHo}{$N_{\rm C^{+};\,H^{0}}$}
\newcommand{\NCpHt}{$N_{\rm C^{+};\,H_{2}}$}
\newcommand{\MCp}{$M_{\rm C^{+}}$}
\newcommand{\NHI}{$N_{\rm HI}$}

\newcommand{\NCNMCII}{$N_{\rm CNM}^{\rm [CII]}$}
\newcommand{\NCNMHI}{$N_{\rm CNM}^{\rm HI}$}
\newcommand{\NHtCO}{$N_{\rm H_{2}}^{\rm CO}$}
\newcommand{\NHtCII}{$N_{\rm H_{2}}^{\rm [CII]}$}
\newcommand{\LCII}{$L_{\rm [CII]}$}
\newcommand{\LCO}{$L_{\rm CO}$}
\newcommand{\fcnm}{$f_{\rm CNM}$}
\newcommand{\class}{\textsc{class}}
\newcommand{\gildas}{\textsc{gildas}}
\newcommand{\herschel}{{\em Herschel}}
\newcommand{\vsys}{$V_{\rm sys}$}
\newcommand{\ergscmsr}{erg~s\per~cm\pers~sr\per}
\newcommand{\Kkms}{\mbox{K~km~s$^{-1}$}}
\newcommand\change{}
\newcommand\changes{}

\newcommand{\AAPF}{\altaffiliation{NSF Astronomy and Astrophysics Postdoctoral Fellow}}
\newcommand{\Arizona}{\affiliation{Steward Observatory, University of Arizona, Tucson, AZ 85721, USA}}
\newcommand{\Maryland}{\affiliation{Department of Astronomy, University of Maryland, College Park, MD 20742, USA}}
\newcommand{\JSI}{\affiliation{Joint Space-Science Institute, University of Maryland, College Park, MD 20742, USA}}
\newcommand{\OSU}{\affiliation{Department of Astronomy, The Ohio State University, Columbus, OH 43210, USA}}
\newcommand{\MPIfR}{\affiliation{Max-Planck-Institut für Radioastronomie, Auf dem Hügel 69, 53121 Bonn, Germany}}
\newcommand{\NRAO}{\affiliation{National Radio Astronomy Observatory, 520 Edgemont Road, Charlottesville, VA 22903}}
\newcommand{\Jansky}{\altaffiliation{Jansky Fellow of the National Radio Astronomy Observatory}}
\newcommand{\Concepcion}{\affiliation{Departamento de Astronom\'ia, Universidad de Concepci\'on, Barrio Universitario, Concepci\'on, Chile}}
\newcommand{\Koln}{\affiliation{I. Physikalisches Institut, Universität zu K\"oln, Z\"ulpicher Str. 77, D-50937 K\"oln, Germany}}
\newcommand{\Kansas}{\affiliation{Department of Physics and Astronomy, University of Kansas, 1251 Wescoe Hall Dr., Lawrence, KS 66045, USA}}
\newcommand{\MPIA}{\affiliation{Max-Planck-Institut f\"ur Astronomie, K\"onigstuhl 17, 69120 Heidelberg, Germany}}
\newcommand{\IPAC}{\affiliation{IPAC, California Institute of Technology, 1200 East California Boulevard, Pasadena, CA 91125, USA}}
\newcommand{\STScI}{\affiliation{Space Telescope Science Institute, 3700 San Martin Drive, Baltimore, MD 21218, USA}}

\shorttitle{\CII\ Spectral Mapping of M\,82}
\shortauthors{Levy et al.}

\begin{document}

\title{\CII\ Spectral Mapping of the Galactic Wind and Starbursting Disk of M82 with SOFIA}

\correspondingauthor{Rebecca C. Levy}
\email{rebeccalevy@arizona.edu}

\author[0000-0003-2508-2586]{Rebecca C. Levy}
\AAPF
\Arizona

\author[0000-0002-5480-5686]{Alberto D. Bolatto}
\Maryland
\JSI

\author[0000-0003-1356-1096 ]{Elizabeth Tarantino}
\STScI
\Maryland

\author[0000-0002-2545-1700]{Adam K. Leroy}
\OSU

\author[0000-0003-3498-2973]{Lee Armus}
\IPAC

\author[0000-0001-6527-6954]{Kimberly L. Emig}
\Jansky
\NRAO

\author[0000-0002-2775-0595]{Rodrigo Herrera-Camus}
\Concepcion

\author[0000-0002-2367-1080]{Daniel P. Marrone}
\Arizona

\author[0000-0001-8782-1992]{Elisabeth Mills}
\Kansas

\author[0000-0002-2155-3259]{Oliver Ricken}
\MPIfR

\author[0000-0001-7658-4397]{Juergen Stutzki}
\Koln

\author[0000-0002-3158-6820]{Sylvain Veilleux}
\Maryland
\JSI

\author[0000-0003-4793-7880]{Fabian Walter}
\MPIA

% ADD YOUR NAME HERE
% Add your affiliation(s) to author_affils.tex
% Add any acknowledgments to acknowledgments section

% \author{Your Name Here}
% \affiliation{TBD}

\begin{abstract}

M\,82 is an archetypal starburst galaxy in the local Universe. The central burst of star formation, thought to be triggered by M\,82’s interaction with other members in the M\,81 group, is driving a multiphase galaxy-scale wind away from the plane of the disk that has been studied across the electromagnetic spectrum. Here, we present new velocity-resolved observations of the [C\textsc{ii}] 158$\mu$m line in the central disk and the southern outflow of M\,82 using the upGREAT instrument onboard SOFIA. We also report the first detections of velocity-resolved ($\Delta V = 10$~km~s$^{-1}$) [C\textsc{ii}] emission in the outflow of M\,82 at projected distances of $\approx1-2$~kpc south of the galaxy center. We compare the [C\textsc{ii}] line profiles to observations of CO and H\textsc{i} and find that \changes{likely} the majority ($>55$\%) of the [C\textsc{ii}] emission in the outflow is associated with the neutral atomic medium. We find that the fraction of [C\textsc{ii}] actually outflowing from M\,82 is small compared to the bulk gas outside the midplane (which may be in a halo or tidal streamers), which has important implications for observations of [C\textsc{ii}] outflows at higher redshift. Finally, by comparing the observed ratio of the [C\textsc{ii}] and CO intensities to models of photodissociation regions, we estimate that the far-ultraviolet (FUV) radiation field in the disk is $\sim10^{3.5}~G_0$, in agreement with previous estimates. In the outflow, however, the FUV radiation field is 2-3 orders of magnitudes lower, which may explain the high fraction of [C\textsc{ii}] arising from the neutral medium in the wind.

\end{abstract}

\section{Introduction}
\label{sec:intro}

Stars inject energy and momentum into their surrounding environments over their entire lifetimes. This so-called stellar feedback can result in massive outflows of gas and dust from the centers of galaxies, where there are high concentrations of stars and star clusters. In these dense environments, energy and momentum injected by supernova explosions and winds from massive stars that are clustered in space and time push material out of the midplane of a galaxy, leading to the observed biconical outflows and superwinds \citep[e.g.,][and references therein]{Heckman1990,Veilleux2005,Veilleux2020}. These outflows are inherently multiphase, with the cool neutral phases of the winds potentially carrying away the bulk of the outflowing mass and the fuel for future star formation \citep[e.g.,][]{Veilleux2020}.

The \CII\ 158\micron\ line in the far-infrared (FIR) is one of the brightest lines in star forming galaxies. It is a major cooling channel of the neutral interstellar medium (ISM) and can contribute 0.1-1\% of the total FIR emission from a galaxy \citep[e.g.,][]{Crawford1985,Stacey1991}. Owing to its low ionization potential (11.2~eV), neutral carbon can be singly ionized in a range of conditions and ISM phases, making the \CII\ 158\micron\ line an excellent tracer of multiphase gas \citep[e.g.,][]{Madden1993,Goldsmith2012,Pineda2013}. In the disks and centers of nearby star forming galaxies, \CII\ tends to be most closely associated with the neutral atomic component, but a substantial fraction can also arise from the molecular component \citep{Mookerjea2016,Rollig2016,Fahrion2017,Tarantino2021}. It is unknown, however, how the origin and distribution of \CII\ may change in a starburst-driven superwind.

M\,82 is an archetypal starburst galaxy, located at a distance of $3.63\pm0.34$~Mpc \citep{Freedman1994} in the M\,81 group. M\,82 is interacting with the other group members \citep{Yun1994,deBlok2018}. It is thought that this tidal interaction triggered a central burst of star formation 10~Myr ago, followed by a second bar-driven burst 5~Myr ago \citep{ForsterSchreiber2003}. The central starburst has launched a multiphase, galaxy-scale wind, which has been studied across the electromagnetic spectrum \citep[e.g.,][]{Lynds1963,Heckman1990,Strickland1997,Walter2002,Engelbracht2006,Strickland2009,Veilleux2009,Yoshida2011,Yamagishi2012,Contursi2013,Beirao2015,Leroy2015,Martini2018,Yoshida2019,Krieger2021}. It is unclear whether the material in the wind has sufficient energy to escape into the intergalactic medium or whether it will fall back onto the galaxy as a fountain \citep[e.g.,][]{Leroy2015,Martini2018,Yuan2023}.

Because M\,82 is such an important laboratory for studying the multiphase ISM in a starburst environment, its center has previously been observed in the \CII\ 158\micron\ line by the Kuiper Airborne Observatory (KAO; \citealt{Stacey1991}) and \herschel\ HIFI and PACS \citep{Loenen2010,Contursi2013,Herrera-Camus2018}. While the PACS data extend into the base of the outflow, \CII\ has not been detected in the outflow at distances greater than $\sim1$~kpc from the midplane. Moreover, because of the low velocity resolution of PACS, the \CII\ data from that instrument are not velocity-resolved.

In this paper, we present new observations of the \CII\ 158\micron\ line taken with the upgraded German REceiver for Astronomy at Terahertz Frequencies (upGREAT; \citealt{Risacher2018}) on board the Stratospheric Observatory for Infrared Astronomy (SOFIA; \citealt{Temi2018}). These velocity-resolved observations measured \CII\ at 10~\kms\ velocity resolution in the inner disk of M\,82 and the southern side of the superwind at distances of $1-2$~kpc from the midplane.

This paper is organized as follows. We describe the data used in this study in Section \ref{sec:obs}. In Section \ref{ssec:results_outflow}, we discuss the main results from the \CII\ detections in the outflow of M\,82. The main results from the \CII\ map in the disk are discussed in Section \ref{ssec:results_disk}. A discussion of the disk and wind together is presented in Section \ref{sec:M82_disk_outflow}.  We discuss constraints on the far-ultraviolet (FUV) radiation field in both the disk and outflow in Section \ref{ssec:radiationfield}. We summarize our main conclusions in Section \ref{sec:summary}.

Throughout, we use CO to refer to $^{12}$C$^{16}$O($J=1-0$). We use $(\alpha,\delta) = (09^{\rm h}55^{\rm m}52.72^{\rm s}, +69^{\circ}40^{\rm m}45.8^{\rm s})$ as the J2000 right ascension and declination of the center of M\,82 \citep{Martini2018}. We assume a central recessional velocity of 210~\kms\ LSRK \citep{Krieger2021}. We assume that the major axis position angle\footnote{The PA is measured counterclockwise from north to the receding side of the galaxy.} (PA) of M\,82 is 67\D; therefore, the PA of the southern side of the wind is 157\D\ \citep[e.g.,][]{Martini2018}. We adopt an inclination of 80\D\ for the disk \citep[e.g.,][]{Lynds1963,McKeith1993,Martini2018,Krieger2021}.  The data and code for the analysis presented here are available online\footnote{\url{https://github.com/rclevy/M82_CII}}.

\section{Observations and Data Reduction}
\label{sec:obs}

\subsection{\CIIsect\ Data from upGREAT}
\label{ssec:obs_sofia}

\begin{figure*}[p]
    \centering
    \includegraphics[width=0.49\textwidth,trim=4.85cm 0cm 7cm 3cm,clip]{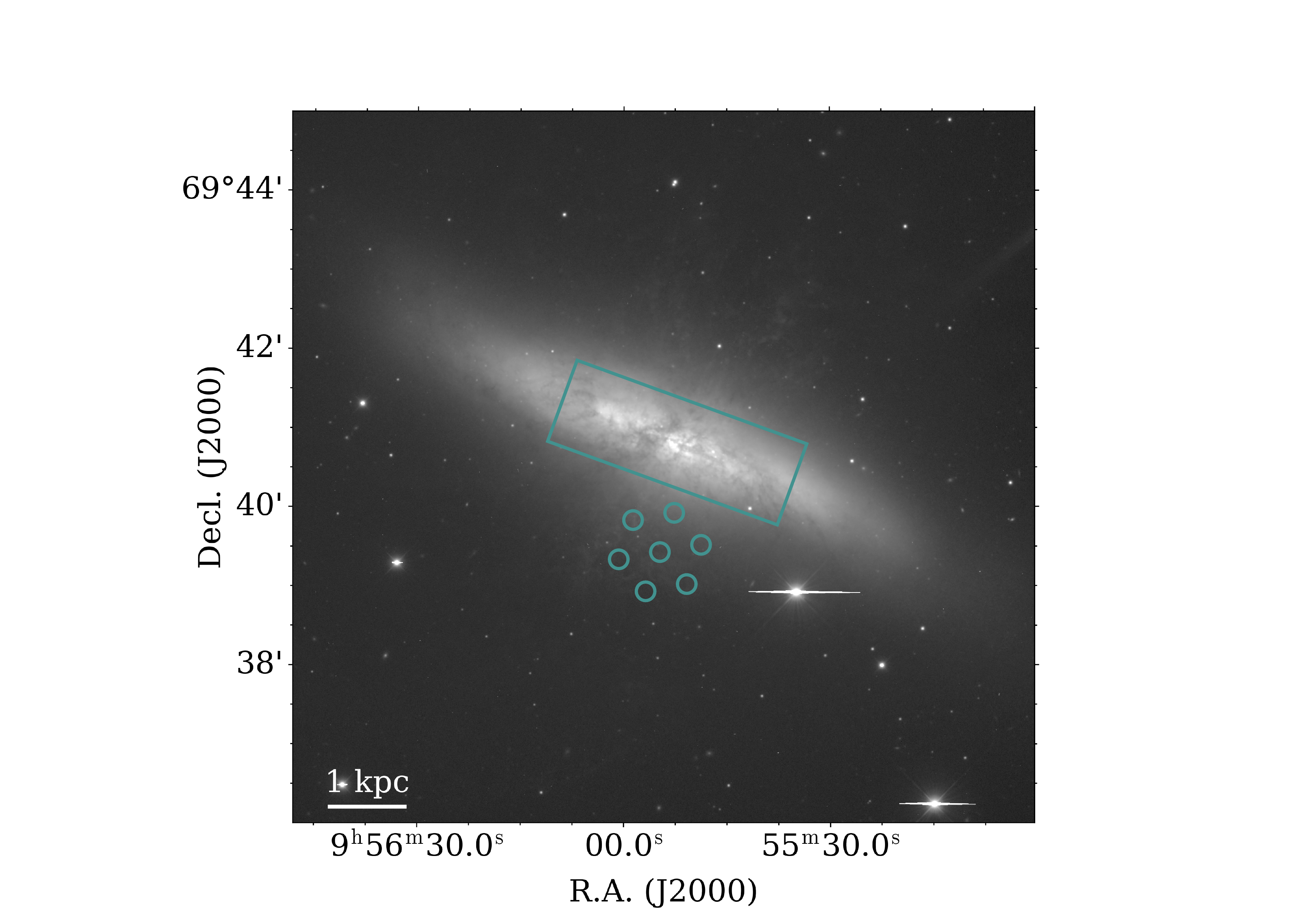}
    \includegraphics[width=0.49\textwidth,trim=4.85cm 0cm 7cm 3cm,clip]{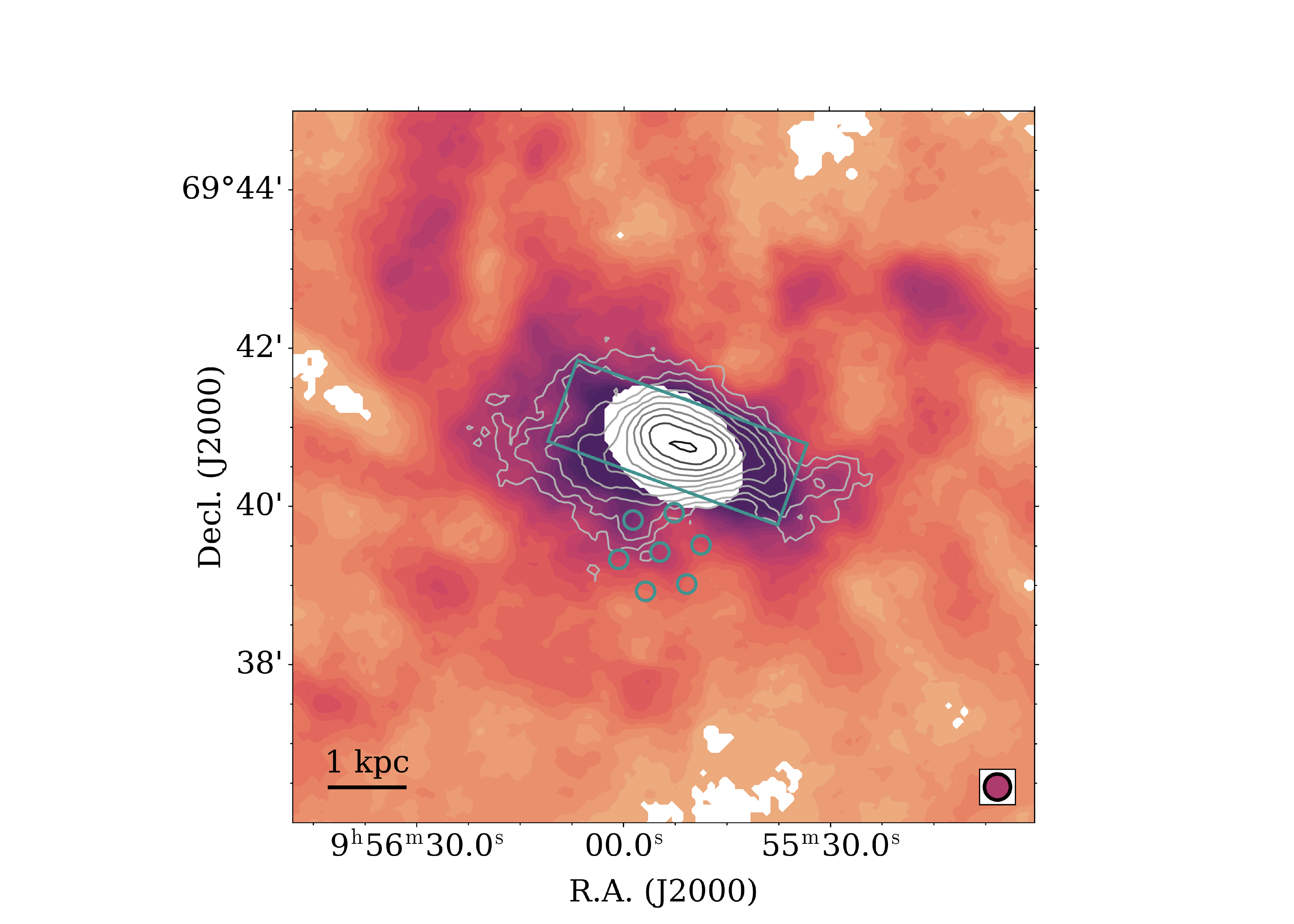}
    \includegraphics[width=0.49\textwidth,trim=3.5cm 0cm 6cm 3cm,clip]{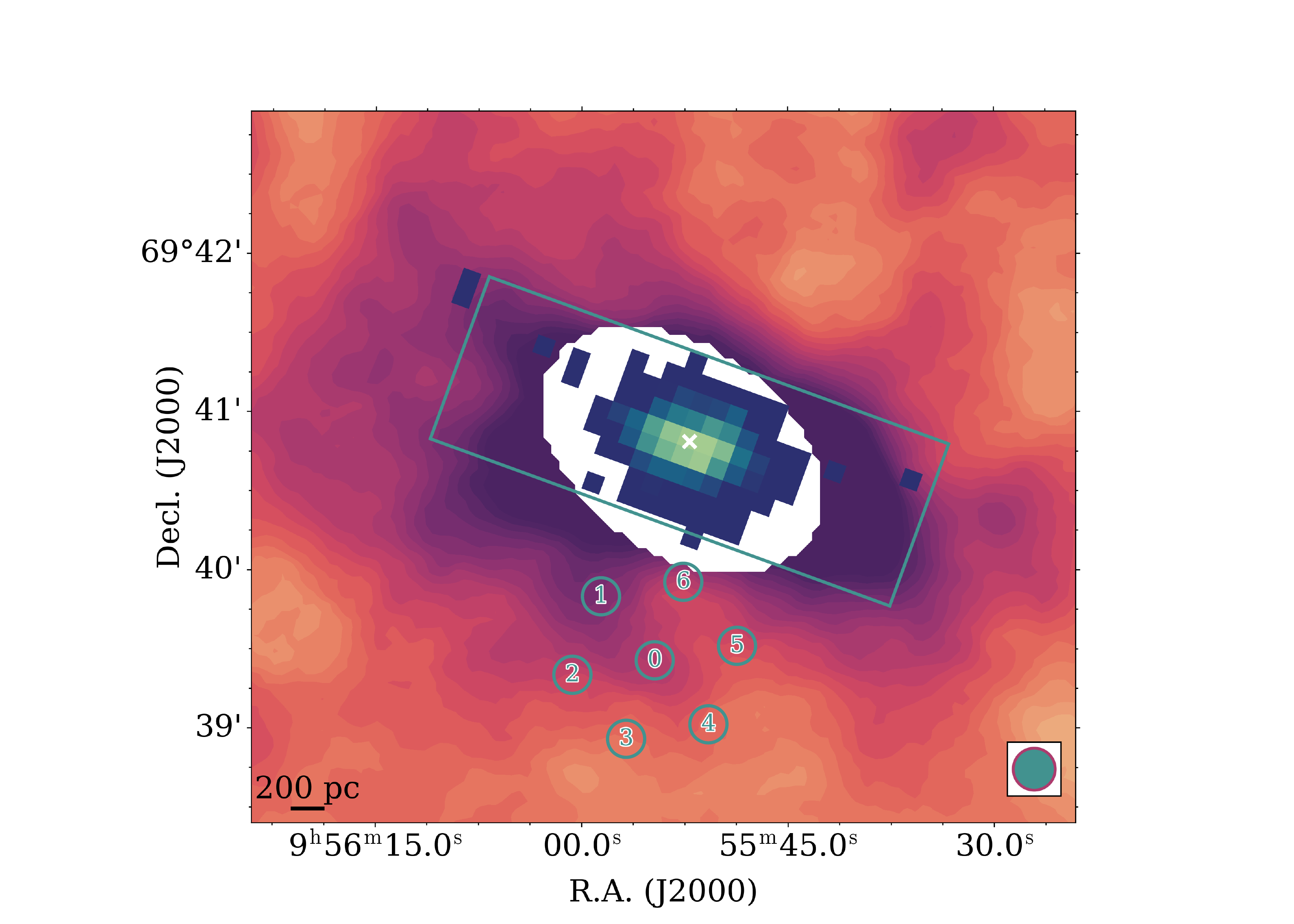}
    \includegraphics[width=0.49\textwidth,trim=3.5cm 0cm 6cm 3cm,clip]{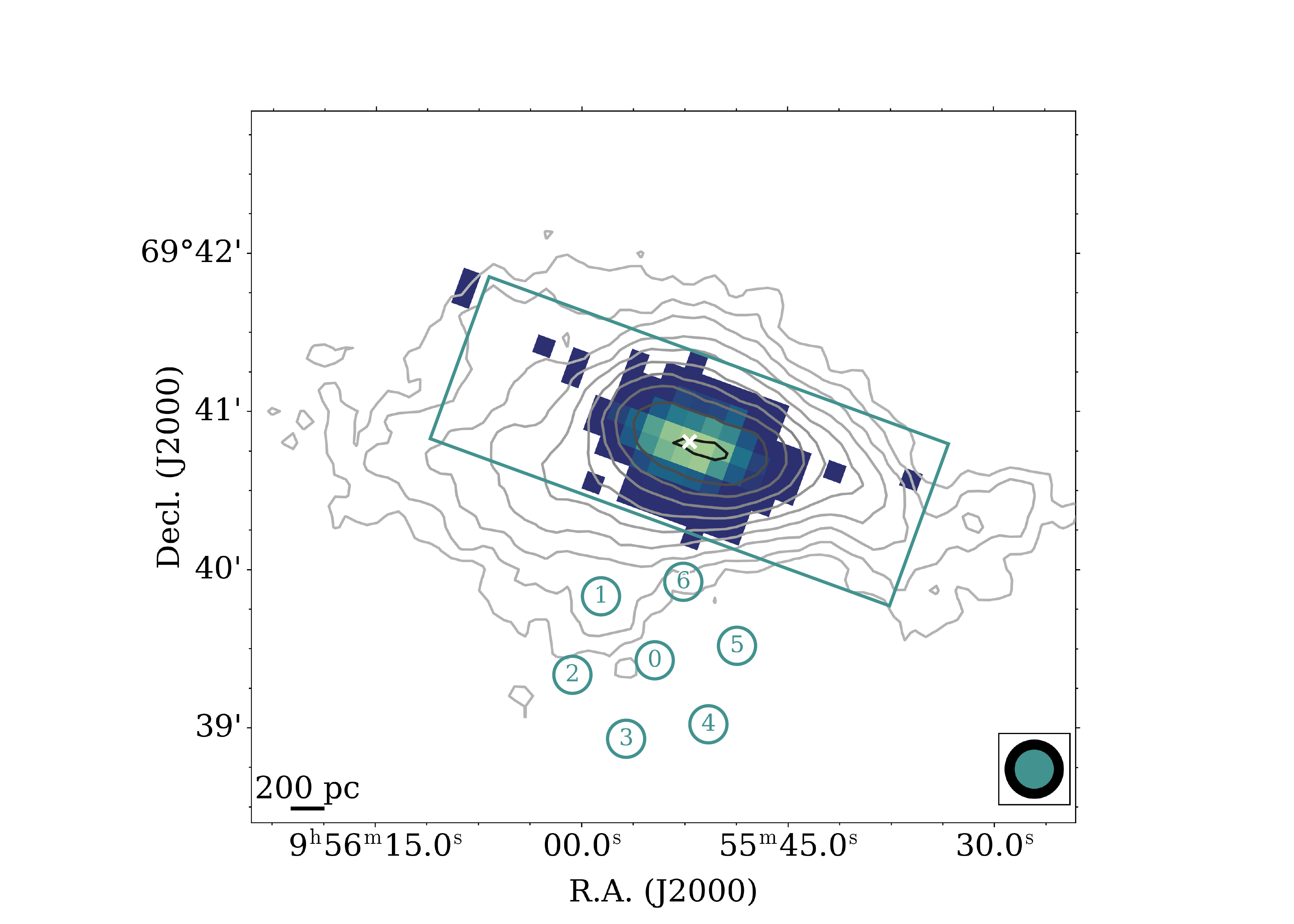}
    \caption{Composite maps of M\,82. In all panels, the teal rectangle shows the fully-sampled region of the \change{upGREAT} \CII\ map of the disk (3~kpc~$\times$~1~kpc). The footprints of the upGREAT single pointing along the southern outflow are shown as the teal circles, where the size of each circle is the upGREAT beam size (14.1\arcsec). Top left: An $r$-band image of M\,82 from the SINGS Survey \citep{Kennicutt2003}.  Top right: The \HI\ integrated intensity is shown in orange-purple filled contours, with values linearly spanning \change{$4-1500$~\Kkms\ \citep{Martini2018}}. The missing \HI\ data in the center reflects where accurate \HI\ information cannot be obtained due to strong absorption against the continuum; this region fully covers the \change{region detected with SNR~$>2$ in the} \CII\ map. The CO integrated intensity is shown in grayscale contours, with values logarithmically spanning $20-600$~K~\kms\ \change{\citep{Krieger2021}}. The FWHM beam sizes of the \HI\ and CO data are shown in the lower right corner in magenta and black respectively. Bottom left: A zoom in, showing the \CII\ integrated intensity in green-blue, with values logarithmically spanning $200-1200$~K~\kms. The \HI\ integrated intensity is shown as the orange-purple filled contours over the same range of values as in the top right panel. The FWHM beam sizes of the \HI\ and \CII\ data are shown in the lower right corner in magenta and teal respectively. The numbers in each of the teal circles give the pixel numbers of the LFA. The white $\times$ marks the galaxy center. Bottom right: The same as the bottom left, but showing the CO integrated intensity (in grayscale contours over the same range of values as in the top right panel) instead of the \HI.}
    \label{fig:composite_map}
\end{figure*}

These observations of the \CII\ 158\micron\ line in M\,82 were taken using the low frequency array (LFA) of the upGREAT\footnote{upGREAT is a development by the MPI f\"ur Radioastronmie and KOSMA/Universit\"at zu K\"oln, in cooperation with the MPI f\"ur Sonnensystemforschung and the DLR Institut f\"ur Optische Systeme.} instrument \citep{Risacher2018} onboard SOFIA \citep{Temi2018}. These data were taken in cycle 8 as part of project 08\_0225 (PI: R. Levy) on 2021 February 19, February 23, February 25, March 10, and March 11.

The upGREAT LFA consists of a seven-pixel hexagonal array for each polarization; the polarizations were averaged for these data. It was tuned to the \CII\ line at 158\micron\ (1.9005~THz). At this frequency, each upGREAT LFA pixel has a half-power beam width of 14.1\arcsec\ ($\approx250$~pc). The bandwidth of the observations ranged from -150$-$430~\kms\ (1.8996$-$1.9033~THz, $\Delta\nu = 3.6$~GHz), with a native velocity resolution of 0.04~\kms\ ($R\approx7.5\times10^6$).

The observing time was split between making an on-the-fly (OTF) map of the inner 3\arcmin$\times$1\arcmin\ of the disk and a single pointing of the upGREAT LFA along the southern outflow. Both observing strategies used a dual-beam-switching mode to measure the OFF positions to maximize baseline stability. More details pertaining to the OTF map and single pointing are given in Sections \ref{ssec:diskmap} and \ref{sssec:outflowpointings}. The footprints of the OTF map and single pointing are shown in Figure \ref{fig:composite_map}. 

All observations were pipeline calibrated (in particular: correction for atmospheric transmission) with the GREAT {\it kalibrate} software \citep{Guan2012} by the upGREAT team and further processed to level-2 data using the \class\ software in \gildas. As part of the calibration, a first-order baseline was removed, the final spectra were smoothed to 10~\kms\ channels, and the spectra were converted to a main-beam temperature (${\rm T_{mb}}$) scale, where $\eta_{\rm mb} \approx\ 0.67$.

\subsubsection{Disk Map}
\label{ssec:diskmap}

To map the inner region of the disk, the LFA was scanned across the central 4.3\arcmin$\times$2.3\arcmin, centered on the galaxy center given in Section \ref{sec:intro}, at an angle of -20\D\ so that the long-axis of the map is aligned with the galaxy's major axis. With this strategy, the fully-sampled region of the map covers the inner 3.1\arcmin$\times$1.1\arcmin\ (3~kpc~$\times$~1~kpc) as shown in Figure \ref{fig:composite_map}. Two tunings (centered at 250~\kms\ and 290~\kms\ LSRK) were used to capture the full extent of the \CII\ emission in the disk. The spectra were weighted by 1/$\sigma_{\rm rms}^2$, where $\sigma_{\rm rms}$ is the root-mean-square (rms) noise of the baseline. The 3D data cube of the disk has an rms noise of 307~mK in 10~\kms\ channels away from the emission. Spatial pixels are 7\arcsec$\times$7\arcsec\ and hence oversample the beam by a factor of $\approx3$ (in area).

\begin{figure*}[p]
    \centering
    \includegraphics[width=0.49\textwidth]{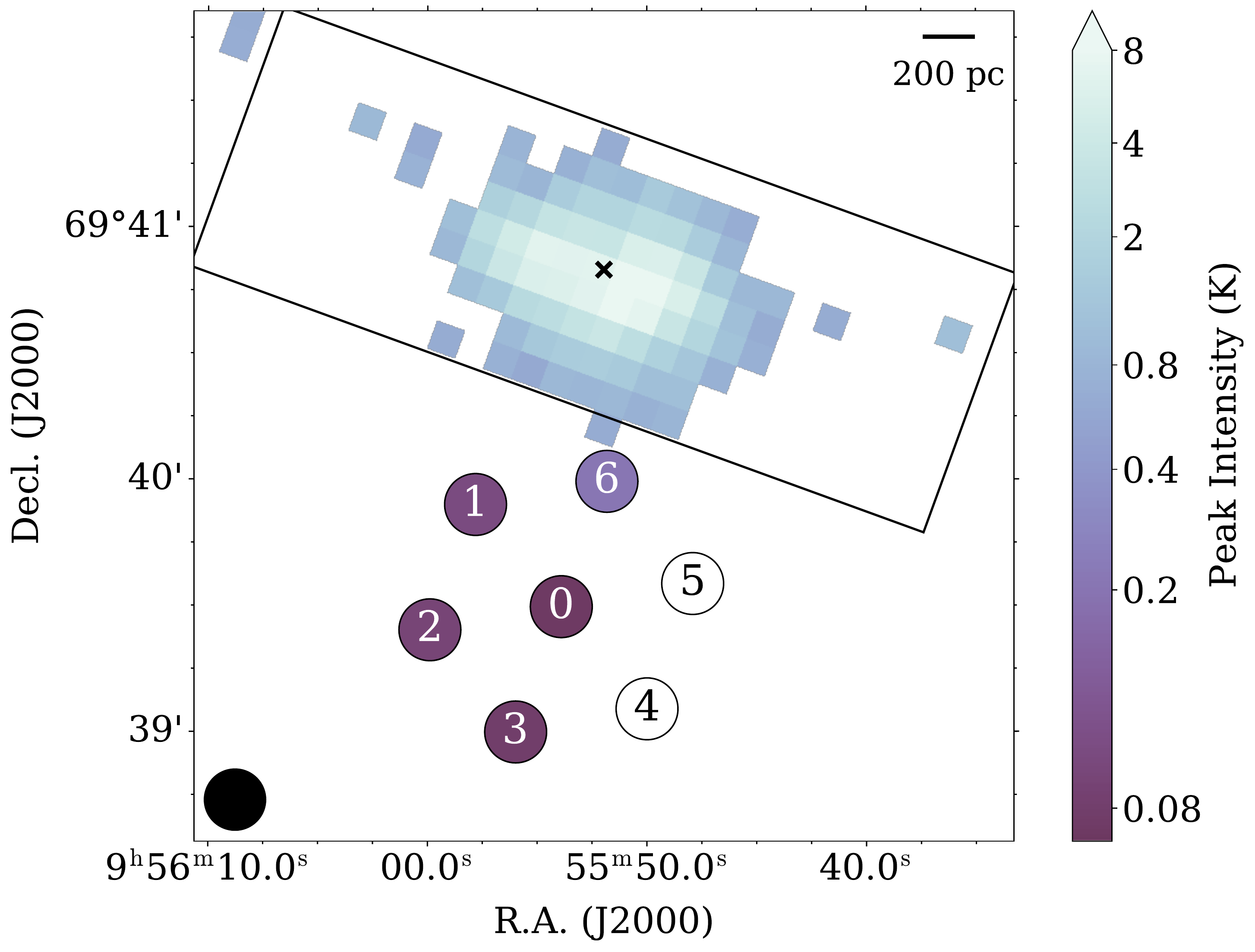}
    \includegraphics[width=0.49\textwidth]{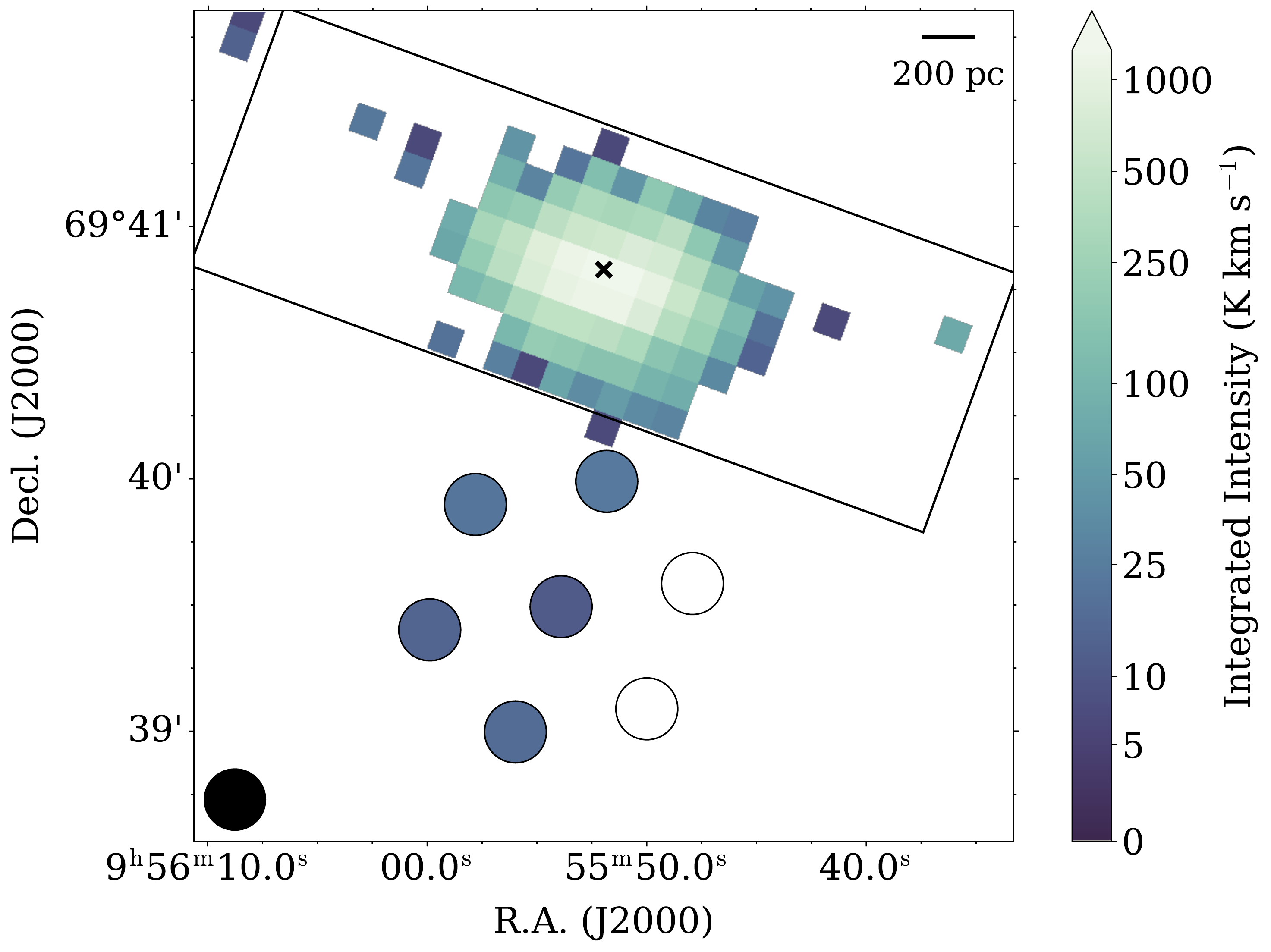}
    \includegraphics[width=0.49\textwidth]{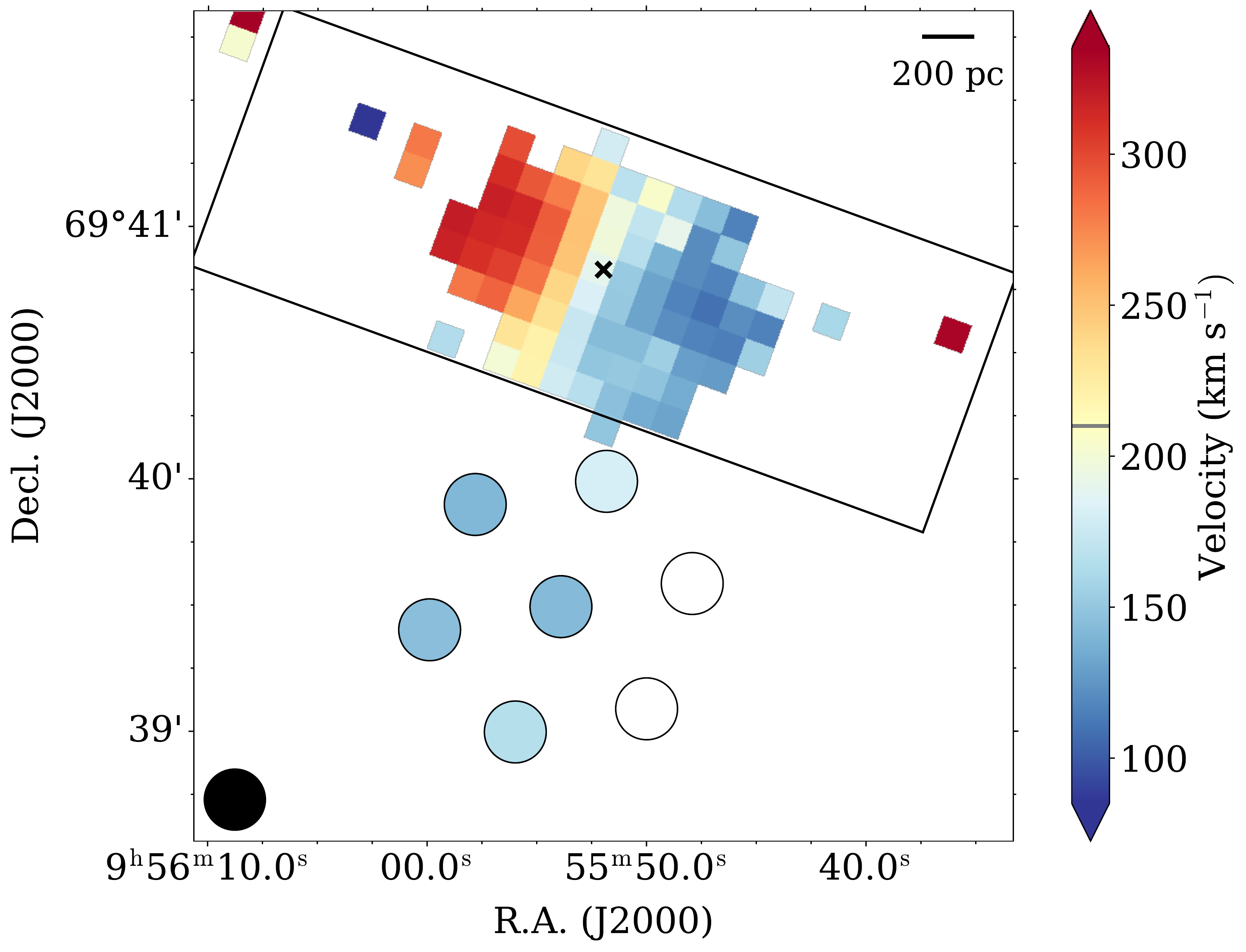}
     \includegraphics[width=0.49\textwidth]{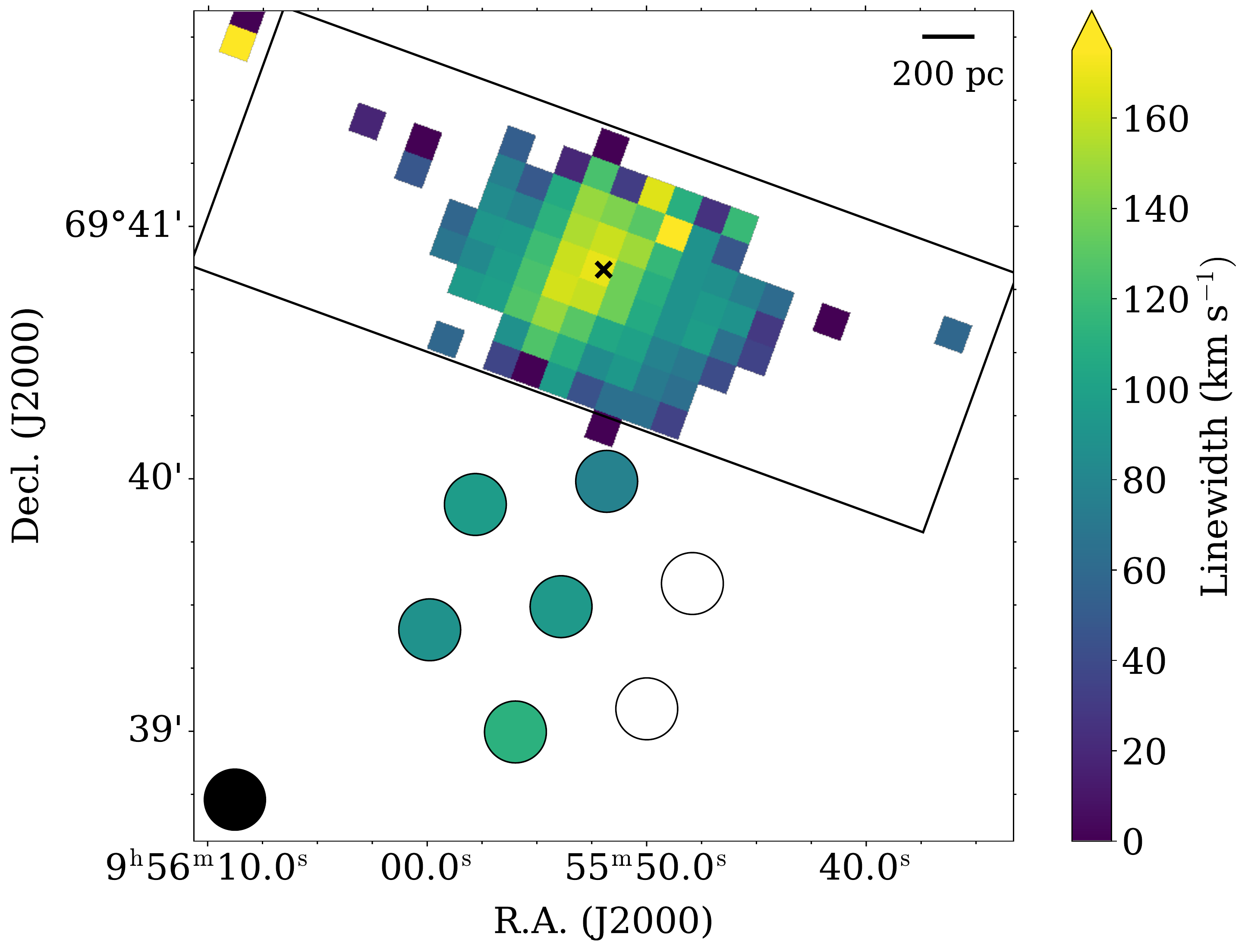}
    \caption{The upGREAT \CII\ maps of the central 3~kpc~$\times$~1~kpc of M\,82, showing the \CII\ peak intensity (top left), integrated intensity (Moment 0; top right), mean velocity (Moment 1; bottom left), and linewidth (Moment 2; bottom right). Spaxels with SNR~$<2$ (relative to the rms noise of the cube) are masked out of the 3D cube before the 2D maps are made. The black rectangle shows the boundaries of the fully-sampled map. The black-outlined circles below the maps show the same values for the single pointings along the outflow. The numbers in each of the circles in the upper left panel give the pixel numbers of the LFA. The solid black circles in the lower left corners show the 14.1\arcsec\ ($\approx250$~pc) beam. The black $\times$ marks the galaxy center \citep{Martini2018}. In the lower left panel, the gray horizontal line in the color bar marks the systemic velocity (\vsys) of M\,82 (210~\kms; \citealt{Krieger2021}).}
    \label{fig:CII_maps}
\end{figure*}

We make 2D maps of the \CII\ peak intensity, integrated intensity, mean velocity, and linewidth using moments\footnote{In both the disk and outflow, the [C{\scriptsize II}] lines are often not Gaussian and so deriving these quantities using a Gaussian fit may lead to biased results.}. We restrict the velocity range to $-100-400$~\kms\ in the calculation of these quantities. For the disk maps, we mask out elements in the cube where the intensity is less than 2$\times$ rms noise ($2\times307 = 614$~mK). These moment maps are shown in Figure \ref{fig:CII_maps}, with the integrated intensity map also being shown in Figure \ref{fig:composite_map}.

As noted in Section \ref{sec:intro}, the \CII\ 158\micron\ line has been previously observed in the center of M\,82 using the KAO \citep{Stacey1991} and \herschel\ PACS \citep{Contursi2013,Herrera-Camus2018} and HIFI \citep{Loenen2010}. \change{In this article, we compare the upGREAT map to that from PACS \citep{Contursi2013}\footnote{\change{The level-2 PACS data were downloaded from the \herschel\ Science Archive, observation ID 1342187205.}}. To robustly compare these measurements, we convolve the PACS cube to the 14\arcsec\ Gaussian beam of upGREAT using the kernels provided by \citet{Aniano2011}. In single beams (not maps), \citet{Tarantino2021} found that not properly accounting for the different point-spread-function (PSF) shapes can lead to 40\% discrepancies between fluxes measured by PACS and upGREAT. We compute the integrated intensity using a moment analysis over the same velocity range as the upGREAT data. We measure the flux from the matched upGREAT and PACS maps in a 1\arcmin~$\times$~1\arcmin\ box (rotated by -20\D) centered on the center of M\,82 \citep{Martini2018}. This box optimizes the overlapping regions of the upGREAT and PACS maps. Over this region, the integrated \CII\ line intensity measured from the upGREAT map is $2.0\times10^{-13}{\rm~W~m^{-2}}$. The matched PACS map yields $1.9\times10^{-13}{\rm~W~m^{-2}}$. These two values agree to within 5\%, well within the 30\% PACS calibration uncertainty \citep{Contursi2013}.}

\subsubsection{Outflow Pointings}
\label{sssec:outflowpointings}

\begin{figure*}[p]
    \centering
    \includegraphics[width=\textwidth]{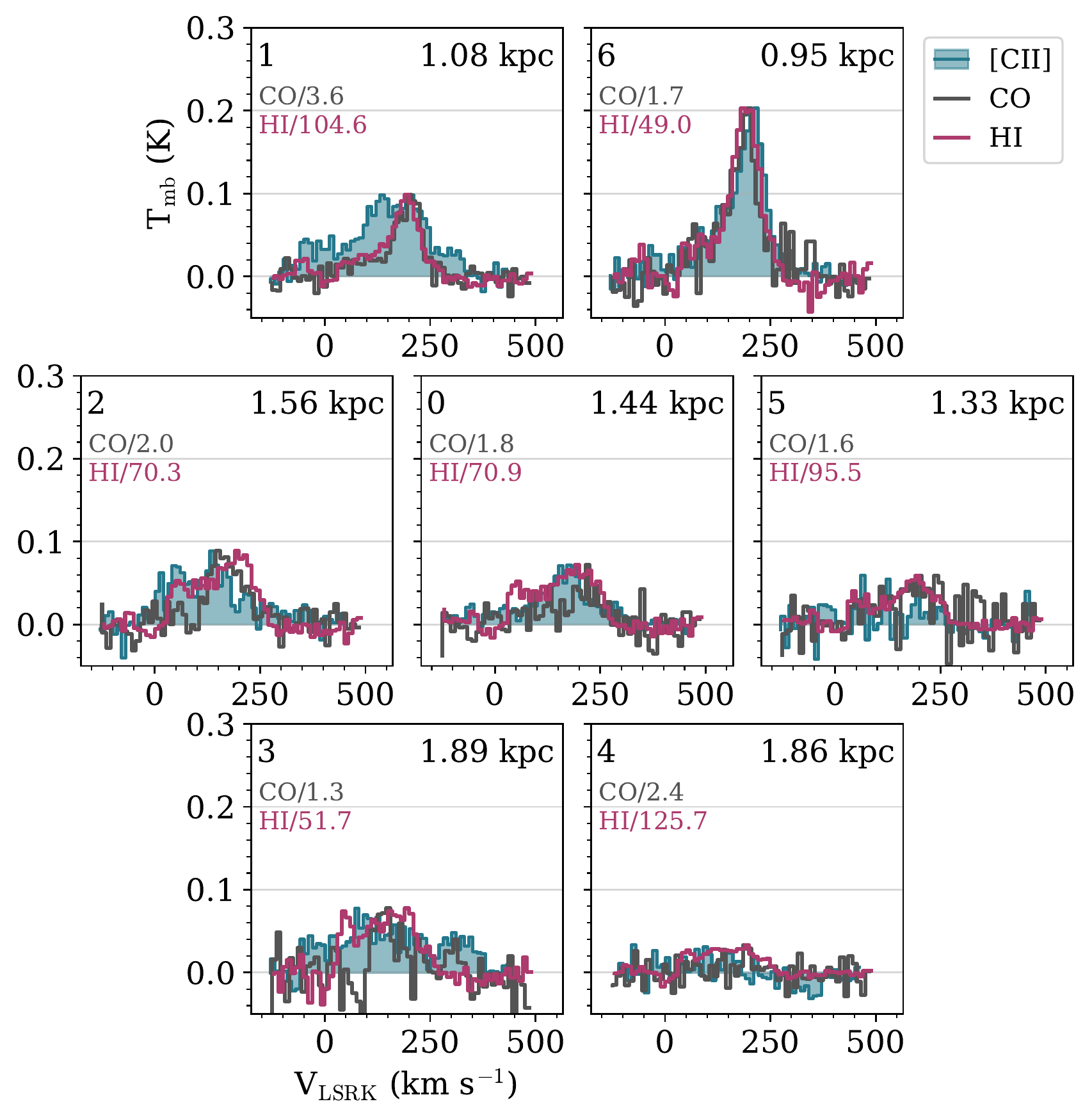}
    \caption{SOFIA upGREAT \CII\ spectra in the southern outflow of M\,82, with 10~\kms\ channels (teal filled). The panels are arranged as shown in Figure \ref{fig:composite_map}, where the pixel number is shown in the top-left of each panel. The distance of the center of each LFA pixel to the galaxy center is shown in the top-right corner. The panels all have the same axis scaling, as shown in the lower-left panel. Scaled versions of the CO (dark gray) and \HI\ (magenta) profiles are overplotted. The CO and \HI\ spectra are simply normalized to the peak of the \CII\ spectrum, and the values by which the CO and \HI\ spectra are scaled are listed in each panel.}
    \label{fig:CII_spectra}
\end{figure*}

To measure \CII\ in the outflow, a single pointing of the upGREAT LFA was used. A location on the southern (brighter) side of the outflow was chosen such that the central pixel of the upGREAT array (Pixel 0) was located 1.5~kpc from the galaxy center along the outflow (i.e., the galaxy minor axis with ${\rm PA} = 157$\D). The LFA array was rotated by 70\D\ to maximize the extent along the southern outflow. The footprint of the LFA is shown is Figures \ref{fig:composite_map} and \ref{fig:CII_maps}. The spectra in the outflow, shown in Figure \ref{fig:CII_spectra}, have rms noise of $7-12$~mK in 10~\kms\ channels.

We calculate moments of the spectra to measure the peak and integrated intensity, the mean velocity, and the linewidth. As with the disk map, we restrict the velocity range to $-100-400$~\kms\ in the calculation of these quantities. We report these values in Table \ref{tab:outflowfits} and show these quantities in Figure \ref{fig:CII_maps}.

\begin{deluxetable*}{cccccccccc}
\tablecaption{Properties of \CII\ spectra in the outflow \label{tab:outflowfits}}
\tablehead{Pixel Number & R.A. & Decl. & ${\rm I_{peak}}$ & ${\rm I_{int}}$ & ${\rm V_0}$ & $\sigma_{\rm V}$ & rms & ${\rm log_{10}M_{C^{+}}}$\\& (J2000 hours) & (J2000 degrees) & (mK) & (K~km~s$^{-1}$) & (km~s$^{-1}$) & (km~s$^{-1}$) & (mK) & (${\rm log_{10}M_\odot}$)}\startdata
0 & $9^{\mathrm{h}}55^{\mathrm{m}}54.6^{\mathrm{s}}$ & $+69^\circ39{}^\prime24.3{}^{\prime\prime}$ & 72.0 $\pm$ 14.8 & 10.8 $\pm$ 1.0 & 143.3 $\pm$ 10.2 & 94.2 $\pm$ 16.9 & 5.8 & 3.5 $\pm 0.2$\\
1 & $9^{\mathrm{h}}55^{\mathrm{m}}56.2^{\mathrm{s}}$ & $+69^\circ39{}^\prime46.9{}^{\prime\prime}$ & 98.6 $\pm$ 33.4 & 21.2 $\pm$ 2.4 & 141.2 $\pm$ 10.3 & 96.0 $\pm$ 19.7 & 9.4 & 3.8 $\pm 0.1$\\
2 & $9^{\mathrm{h}}55^{\mathrm{m}}56.9^{\mathrm{s}}$ & $+69^\circ39{}^\prime19.8{}^{\prime\prime}$ & 88.9 $\pm$ 28.2 & 13.7 $\pm$ 2.0 & 144.9 $\pm$ 10.5 & 89.3 $\pm$ 17.5 & 9.6 & 3.6 $\pm 0.2$\\
3 & $9^{\mathrm{h}}55^{\mathrm{m}}55.3^{\mathrm{s}}$ & $+69^\circ38{}^\prime59.1{}^{\prime\prime}$ & 77.5 $\pm$ 31.0 & 16.4 $\pm$ 2.2 & 165.1 $\pm$ 10.6 & 111.8 $\pm$ 93.8 & 6.9 & 3.7 $\pm 0.1$\\
4 & $9^{\mathrm{h}}55^{\mathrm{m}}53.4^{\mathrm{s}}$ & $+69^\circ39{}^\prime00.0{}^{\prime\prime}$ & --- & --- & --- & --- & 6.8 & ---\\
5 & $9^{\mathrm{h}}55^{\mathrm{m}}52.5^{\mathrm{s}}$ & $+69^\circ39{}^\prime30.1{}^{\prime\prime}$ & --- & --- & --- & --- & 16.8 & ---\\
6 & $9^{\mathrm{h}}55^{\mathrm{m}}54.1^{\mathrm{s}}$ & $+69^\circ39{}^\prime52.3{}^{\prime\prime}$ & 203.0 $\pm$ 27.4 & 23.3 $\pm$ 1.9 & 180.2 $\pm$ 10.3 & 77.4 $\pm$ 247.0 & 11.5 & 3.8 $\pm 0.1$\\
\enddata
\tablecomments{The R.A. and Decl. of the center of each pixel are given. The other columns show the moments of the spectra including the peak intensity (${\rm I_{peak}}$), the integrated intensity (moment 0; ${\rm I_{int}}$), the mean velocity (moment 1; ${\rm V_0}$), the linewidth (moment 2, $\sigma_{\rm V}$), and corresponding uncertainties. The moments are calculated over a fixed velocity range spanning -$100-400$~\kms. rms is the root-mean-square noise of the spectrum calculated outside of the velocity range used for the moments. If the moments cannot be calculated because the line is not detected, then the rms is calculated over the entire bandpass. All temperatures refer to T$_{\rm mb}$. ${\rm M_{C^{+}}}$ is the mass in C$^+$ inferred from the spectrum considering only collisions with the atomic gas; see Appendix \ref{app:density_calc} for details of this calculation.} 

\end{deluxetable*}

\change{As a check, we compare the flux we measure in outflow Pixels 1 and 6 to the \herschel\ PACS map \citep{Contursi2013} convolved to a 14\arcsec\ Gaussian PSF as described in Section \ref{ssec:diskmap}. Pixels 1 and 6 are the only upGREAT pointings that overlap with the PACS coverage. From the PACS map, we measure a total \CII\ integrated line intensity of 4.8~\Kkms\ (12.0~\Kkms) in Pixel 1 (6). This is lower than the \CII\ integrated intensity of 21.2~\Kkms\ (23.3~\Kkms) measured in the upGREAT map (Table \ref{tab:outflowfits}). We note, however, that these pixels are at the edges of the PACS map and there may be substantial flux loss due to edge effects and undersampling. When these PACS maps were reprocessed by \citet{Herrera-Camus2018}, for example, these edge regions were excluded.}

\subsection{Ancillary Data}
\label{ssec:obs_co_hi}

Because \CII\ can be excited in many conditions, we compare the \CII\ emission with the molecular and atomic ISM components. We use ancillary CO and \HI\ data as tracers of those respective components.

\subsubsection{CO(1-0) Tracing the Molecular Gas}
The CO data used in this study is from the Institut de Radioastronomie Millim\'etrique (IRAM) 30-m telescope. These data were presented by \citet{Krieger2021}, and we direct the reader to that paper for details on the observations, calibration, and imaging. We note that while the data presented by \citet{Krieger2021} include both interferometric and single-dish observations, we find that there are significant interferometric artifacts (absorption) in the spectra that hinder the comparison to the \CII\ data in this work. Therefore, we use only the single-dish data, which do not show these spectral artifacts. The final CO cube has a spatial resolution of 22\arcsec\ ($\approx385$~pc) and a velocity resolution of 5~\kms. For this work, we smooth the CO cube to a velocity resolution of 10~\kms\ to match the \CII\ data. We show the CO integrated intensity in Figure~\ref{fig:composite_map}.

\subsubsection{\HIsect\ Tracing the Atomic Gas}
\label{sssec:HI}
The \HI\ data used in this study is from the Karl G. Jansky Very Large Array (VLA) \change{which has been combined with single-dish data from the Robert C. Byrd Green Bank Telescope (GBT)}. These data were presented by \citet{Martini2018}, and we direct the reader to that paper for details on the observations, calibration, and imaging. \change{The \HI\ cube has a native velocity resolution of 5~\kms.} We spectrally smooth this data to a velocity resolution of 10~\kms\ to match the \CII\ data. \change{The \HI\ data used here have a spatial resolution of 17\arcsec\ ($\approx300$~pc), slightly higher than that presented by \citet{Martini2018}}. In the center of M\,82, the \HI\ is heavily absorbed against the bright continuum \citep{Martini2018}, so the \HI\ cannot be compared to the \CII\ in the disk; this is not a problem in the outflow pointings. To produce the integrated intensity map (Figure \ref{fig:composite_map}), we create a mask to exclude the absorption and to exclude channels where the \HI\ intensity $<2.3$~K (SNR$<3$). We note that our analysis is performed on the \HI\ cube itself, and the \HI\ integrated intensity map is to aid in the visualization.

\subsubsection{$r$-band Image}
For visualization purposes, we show an $r$-band image of M\,82 in Figure \ref{fig:composite_map}. This image was taken with as part of the SINGS Survey \change{\citep{Kennicutt2003,SINGS}} and was downloaded from the NASA/IPAC Extragalactic Database.

\subsubsection{Matching the Datasets}
In the outflow, we extract the CO and \HI\ from the central pixel of each upGREAT pointing. As noted above, the CO and \HI\ data are smoothed to a velocity resolution of 10~\kms, to match the \CII. To make the most accurate comparison, the data should be convolved to the 22\arcsec\ resolution of the CO data. Because, however, the upGREAT data in the outflow are single pointings (not a map), they cannot be convolved to lower resolution. Therefore, we do not match the beam sizes of the CO, \HI, and \CII\ data in the outflow. As we will discuss in Section \ref{ssec:results_disk}, we do match the resolutions and pixel scales of the \CII\ and CO datasets in the disk.

We show the \CII, CO, and \HI\ spectra in Figure \ref{fig:CII_spectra}, where the CO and \HI\ are simply normalized to the peak intensity of the \CII. We derive the CO and \HI\ integrated intensities (moment 0) in the same way as for the \CII\ spectra and over the same velocity range. In Table \ref{tab:outflowfits_HICO}, we give the ratios of the integrated intensities of \HI/\CII\ and CO/\CII, where all integrated intensities are in \Kkms\ units.

\begin{deluxetable}{cccc}
\tablecaption{Ratios of the integrated intensities and luminosities in the outflow.\label{tab:outflowfits_HICO}}
\tablehead{Pixel Number & ${\rm I_{int,HI}/I_{int,[CII]}}$ & ${\rm I_{int,CO}/I_{int,[CII]}}$ & ${\rm L_{[CII]}/L_{CO}}$}
\startdata
0 & 1372.9 $\pm$ 140.5 & 12.0 $\pm$ 1.3 & 154 $\pm$ 6\\
1 & 779.1 $\pm$ 91.1 & 12.0 $\pm$ 1.4 & 154 $\pm$ 6\\
2 & 944.4 $\pm$ 141.2 & 11.4 $\pm$ 1.7 & 163 $\pm$ 7\\
3 & 498.5 $\pm$ 69.5 & 3.5 $\pm$ 0.5 & 535 $\pm$ 30\\
4 & --- & --- & ---\\
5 & --- & --- & ---\\
6 & 718.3 $\pm$ 65.7 & 14.8 $\pm$ 1.3 & 125 $\pm$ 5\\
\enddata
\tablecomments{\change{To form the ratios, all integrated intensities are in K~km~s$^{-1}$ units and all luminosities are in $L_\odot$ units.}} 

\end{deluxetable}

\section{\CIIsect\ in the Wind of M\,82}
\label{ssec:results_outflow}

We robustly detect the \CII\ 158\micron\ line in five of the seven LFA pixels in the southern outflow of M\,82, as shown in Figure \ref{fig:CII_spectra}. As described in Section \ref{sssec:outflowpointings}, we calculate moments of these spectra which are listed in Table \ref{tab:outflowfits} and shown in Figure \ref{fig:CII_maps}. The ratios of the peak CO to \CII\ intensities (in K brightness temperature units) are $\sim1$, in agreement with previous work \citep{Stacey1991}. We present the integrated intensity ratios (on a \Kkms\ intensity scale) in each LFA pixel in Table \ref{tab:outflowfits_HICO}.

From the \CII\ spectra in the outflow, we calculate the column density and mass of \Cp\ in each outflow pointing. These calculations are detailed in Appendix \ref{app:density_calc}. Briefly, the \Cp\ column density as a function of velocity (\NCp) is calculated the following  Equation \ref{eq:crawford_simp}, assuming that the \CII\ is only excited in the \change{cold neutral medium \citep[CNM; e.g.,][]{Pineda2013,Fahrion2017,Herrera-Camus2017}}. For this calculation, we must assume a kinetic temperature, $T$, and a gas density, $n$. Our assumed temperature and density come from inspecting the spatially-resolved photodissociation region (PDR) modeling results presented by \citet{Contursi2013}. Though they do not probe out as far into the outflow as our measurements, they find temperature of $T\sim200-300$~K and densities of $n\sim10-150$~cm\pert\ along the southern outflow away from the disk (uncorrected for the effects of the ionized gas; see their Figures 15 and 16). We note that these values differ from those presented in their Table 1. The "southern outflow" macro region they define is likely contaminated by the starburst (see their Figure 18) and hence the average temperature and density reported in that table are likely too high to apply to the part of the outflow we are studying. From these results, we take a representative temperature of $T=250\pm50$~K and a representative density of $n=100\pm50$~cm\pert. 

As we discuss in more detail below, our choice to assume that the \CII\ is primarily excited in the CNM is well-justified. We sum this column density over velocity and multiply by the area of each LFA pixel to measure the mass in \Cp\ (\MCp; Equation \ref{eq:MCp}). We list \MCp\ in each pointing in Table \ref{tab:outflowfits}. The total \Cp\ mass in the part of the outflow covered by these pointings is $2.4\times10^4$~\msun\ (excluding Pixels 4 and 5 where the \CII\ line is not detected).

\subsection{Atomic Gas in the Outflow}
\label{ssec:atomgas_outflow}
\begin{figure*}
    \centering
    \includegraphics[width=\textwidth]{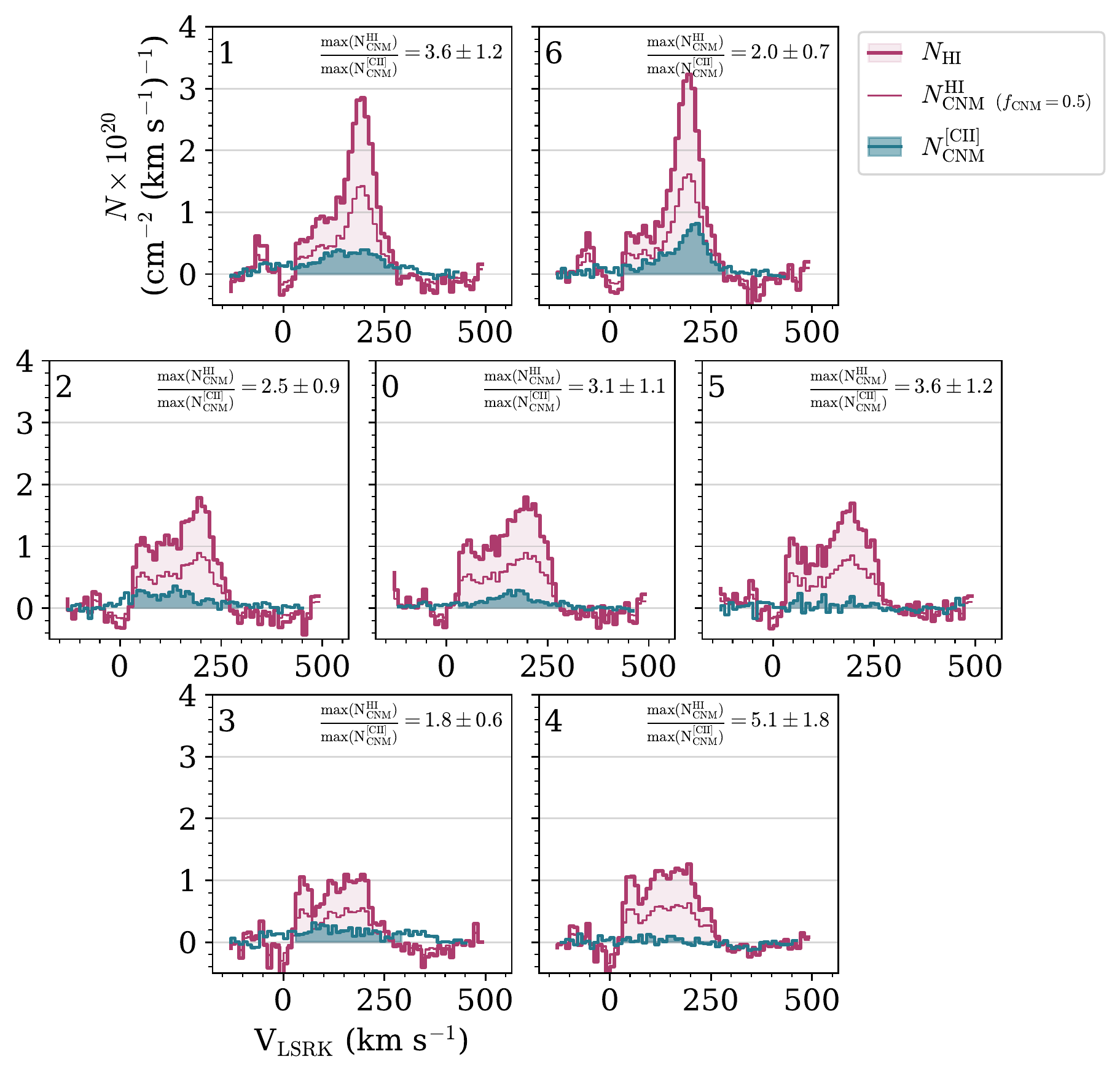}
    \caption{The \HI\ and CNM column densities (per unit velocity) in the outflow of M\,82. The teal curves show the CNM column density inferred from the \CII\ spectrum (\NCNMCII) assuming collisions with only the atomic gas, \change{$T = 250$~K, $n = 100$~cm$^{-3}$,} and ${\rm C/H = 1.5\times10^{-4}}$. %In the high $T_{\rm CNM}$ and $n_{\rm CNM}$ limit, the maximum \NCNMCII\ is $\sim10^{18}$~cm\pers~(\kms)\per, much lower than the measured \NCNMCII.
    The thick magenta curves show \NHI\ measured directly from the \HI\ data. The thin magenta curves show the CNM column density inferred from the \HI\ (\NCNMHI) assuming \fcnm~=0.5. The LFA pixels are oriented as in Figure \ref{fig:CII_spectra}, and the pixel numbers are given in the top left corner of each panel. The number in the top right corner of each panel is the ratio of the peaks of \NCNMHI\ and \NCNMCII\ over the shaded portions of the profiles. The uncertainties come from a Monte Carlo over the range of temperature and densities considered. }
    \label{fig:columndensity_H0}
\end{figure*}

From \NCp, we can estimate the effective CNM column density needed to produce the observed \CII\ spectra (see details in Section \ref{app:atomic}). To make this conversion, we divide \NCp\ by a C/H abundance ratio (${\rm C/H} = 1.5\times10^{-4}$; \citealt{Gerin2015}). We show these effective CNM column density profiles (per unit velocity, i.e., divided by the channel width of 10~\kms) based on the \CII\ (\NCNMCII) in teal in Figure~\ref{fig:columndensity_H0}.

\begin{figure*}
    \centering
    \includegraphics[width=\textwidth]{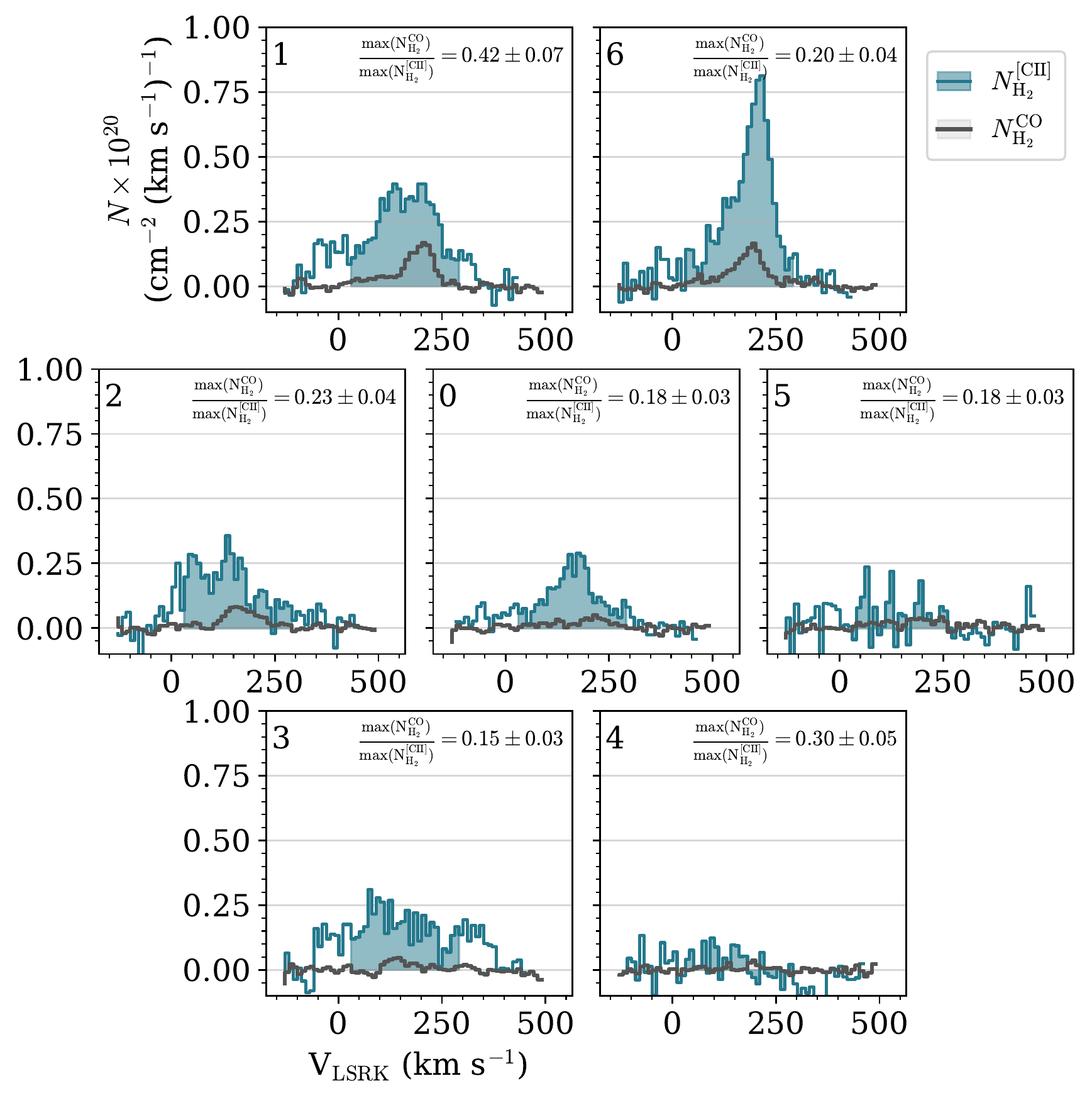}
    \caption{Similar to Figure \ref{fig:columndensity_H0}, but showing the \htwo\ column densities (per unit velocity) in the outflow of M\,82. The teal curves show the \htwo\ column density inferred from the \CII\ spectrum (\NHtCII) assuming collisions with only \htwo, \change{$T = 250$~K, $n = 100$~cm$^{-3}$,} and \change{${\rm C/H_2 = 3\times10^{-4}}$.} The gray curves show the column density of \htwo\ measured from the CO data (\NHtCO) assuming $X_{\rm CO}=0.5\times10^{20}$~cm\pers~(K~\kms)\per. The number in the top right corner of each panel is the ratio of the peaks of \NHtCO\ and \NHtCII\ over the shaded portions of the profiles. The uncertainties come from a Monte Carlo over the range of temperature and densities considered.}
    \label{fig:columndensity_H2}
\end{figure*}

In order to determine the ISM phase from which most of the \CII\ emission originates, we compare our \NCNMCII\ profiles to the measured total \HI\ column density profiles (\NHI). Modulo optical depth effects, the \HI\ profiles (measured directly from the \HI\ data) probe the total atomic gas content from both the CNM and WNM. Following \citet{Martini2018}, we convert the \HI\ intensity to a column density where 
\begin{equation}
    \label{eq:HI_int2column}
    N_{\rm HI} = 1.823\times10^{18}\left[\frac{I_{\rm HI}}{\rm K~km~s^{-1}}\right]~{\rm cm~s^{-2}}
\end{equation}
which assumes optically thin emission. These column density profiles (per unit velocity) are shown in magenta in Figure \ref{fig:columndensity_H0}. For all LFA pixels, \NHI~$>$~\NCNMCII, meaning there is sufficient \HI\ to fully explain the \CII\ emission without the need to invoke optical depth effects. 

A somewhat more direct comparison of the column densities can be made by assuming some CNM fraction (\fcnm) of the atomic component, since the \CII\ emission is thought to only arise from the cold phase \citep[e.g.,][]{Pineda2013,Fahrion2017,Herrera-Camus2017}. In Figure \ref{fig:columndensity_H0}, we show \NCNMHI~$\equiv$\fcnm\NHI\ for \fcnm~$=0.5$ (thin magenta lines). In all LFA pixels, \NCNMHI~$\gtrsim$~\NCNMCII, meaning that there is enough CNM to fully explain all of the \CII\ emission. The ratio of \NCNMHI\ to \NCNMCII\ at the peak of the \HI\ profile is given in the upper right of each panel in Figure \ref{fig:columndensity_H0}. \change{By varying \fcnm, we find that the \CII\ emission can be fully explained by collisions with the atomic gas for \fcnm~$>0.2$.}

%\change{We can use this comparison to place a lower limit on \fcnm\ in the outflow of M\,82. Smaller values of \fcnm\ result in lower \NCNMHI\ (thin magenta curve in Figure \ref{fig:columndensity_H0}). If \NCNMHI~$<$~\NCNMCII, this indicates that there is insufficient cold atomic gas to fully explain the observed \CII\ by collisions with H$^0$ alone. For these spectra, \NCNMHI~$\approx$~\NCNMCII\ for \fcnm~$\approx 0.2$ (i.e., 20\% (80\%) of the total \HI\ is in the CNM (WNM)). As discussed in the following sections, $<40\%$ of the \CII\ arises from collisions with molecular or ionized gas, meaning that $>60\%$ of the \CII\ emission must arise from collisions with the atomic gas.}

%Said another way, the CNM fraction in the outflow must be $\gtrsim0.5$. \CII\ emission can arise from the atomic, molecular (including CO-dark), and ionized phases. At smaller values of \fcnm, the atomic component fully explains the \CII\ profiles, with no room for contributions from the molecular or ionized gas phases (described below). Therefore, the atomic component of the outflow appears to be more dominated by a cold phase\footnote{We note that this is at odds with the assumptions made by \citet{Yuan2023}. In their model of the outflow of M\,82, they assume that all of the atomic material is warm and optically thin.}.

To assess the impact of uncertainties in our assumed temperature and density, we perform a Monte Carlo over both quantities. Using 500 trials, we allow both $T$ and $n$ to vary uniformly over the uncertainty range when calculating the column density. We define the uncertainty on the column density as the standard deviation of the trials. We propagate this uncertainty through to the ratio shown in the upper right of each panel in Figure \ref{fig:columndensity_H0}. %Therefore, even with the uncertainties on $T$ and $n$, the above conclusions that there is enough CNM to fully explain the \CII\ emission and that \fcnm~$\gtrsim0.5$ hold. 

%That said, t
\change{There is evidence of a warm atomic phase in some of the spectra.} Focusing specifically on Pixel 1, the \CII\ profile is more "flat-topped" than the \HI\ (both in Figure \ref{fig:CII_spectra} and Figure \ref{fig:columndensity_H0}). This peak in the \HI\ profile around ${\rm V_{LSRK}}=200$~\kms\ may be indicative of an appreciable WNM component. A similar WNM component is seen in Pixel 2 at a similar velocity. In Pixel 6, however, the atomic gas appears to be dominated by a cold component. Thus there are variations in the overall density and temperature of the atomic component within the outflow of M\,82, likely because the entrained material is clumpy.

\subsection{Molecular Gas in the Outflow}
\label{ssec:molgas_outflow}
As the \CII\ emission can also arise from the molecular phase, we also calculate \NCp\ assuming collisions with only \htwo\ (see details in Section \ref{app:mol}). We assume the same temperature and density as for collisions with the atomic gas and \change{we assume an abundance ratio of ${\rm C/H_2=2\times C/H =3.0\times10^{-4}}$.} We show these effective \htwo\ column density profiles (per unit velocity) based on the \CII\ (\NHtCII) in teal in Figure \ref{fig:columndensity_H2}. We repeat the Monte Carlo analysis over the uncertainties in $T$ and $n$ as above.

To compare, we calculate the column density of \htwo\ from the CO data (\NHtCO). To convert from the CO intensity to column density (per unit velocity), we assume a "starburst" CO-to-\htwo\ conversion factor $X_{\rm CO}=0.5\times10^{20}$~cm\pers~(K~\kms)\per\ \citep[e.g.,][]{Bolatto2013,Krieger2021}. The value is slightly smaller than the single value of $X_{\rm CO}=0.7\times10^{20}$~cm\pers~(K~\kms)\per\ used by \citet{Leroy2015}, where we have converted from CO(2-1) to CO(1-0) assuming $I_{\rm CO~2-1}/I_{\rm CO~1-0}=0.7$. In their model of M\,82, \citet{Yuan2023} find $X_{\rm CO}\approx0.2\times10^{20}$~cm\pers~(K~\kms)\per\ for this region of the southern outflow\footnote{In their Figure 13, \citet{Yuan2023} report $X_{\rm CO~2-1}$. We have converted this to $X_{\rm CO~1-0}$ assuming $I_{\rm CO~2-1}/I_{\rm CO~1-0}=0.7$ \citep{Leroy2015}.}.

We show these molecular gas column density profiles in gray in Figure \ref{fig:columndensity_H2}, assuming our fiducial $X_{\rm CO}=0.5\times10^{20}$~cm\pers~(K~\kms)\per. We note that adopting the conversion factors used by \citet{Leroy2015} or \cite{Yuan2023} produce negligible changes in the \NHtCO\ profiles. In all LFA pixels, \NHtCO~$\sim15-40$\%~\NHtCII, with most pixels $\sim20\%$, meaning that the molecular gas (traced by CO) is not the dominant contributor to the overall \CII\ emission. %It is possible that there is CO-dark molecular gas in the outflow and that this may contribute to the \CII\ emission, but it is unlikely to fully explain the difference.
In summary, because only $\sim20$\% of the \CII\ can be excited by the molecular ISM, we conclude that the majority of the \CII\ arises from the CNM. \change{This result has important connections to studies of \CII\ in dust \change{star-forming} galaxies at higher redshifts, where the \CII\ 158\micron\ line is visible with interferometers such as ALMA. While it is tempting to use the \CII\ lines as a tracer of molecular gas and/or hence star formation rate \citep[e.g.,][]{Herrera-Camus2015,Zanella2018,Dessauges-Zavadsky2020}, caution should be used as these results in M\,82 suggest that the majority of the \CII\ does not arise from the molecular, star-forming material.}

\subsubsection{\change{CO-dark Molecular Gas}}
\change{CO does not perfectly trace the total molecular gas content of the ISM, and this component of missed gas is called CO-dark molecular gas \citep[e.g.,][]{Grenier2005,Langer2010,Wolfire2010}. By its nature, this phase is difficult to study. In the Milky Way, the fraction of molecular gas not traced by CO varies with galactocentric radius and cloud density, dropping below 20\% within 4~kpc of the center and in dense clouds, but reaching nearly 80\% in the diffuse ISM and beyond 10~kpc \citep{Pineda2013,Langer2014}. From their models, \citet{Wolfire2010} found that the fraction of CO-dark gas in PDRs is relatively constant with the ambient radiation field and that the main driver is the cloud's visual extinction, $A_V$ --- a measure of the dust shielding --- where the fraction of CO-dark molecular gas decreases steeply with increasing $A_V$. The fraction of CO-dark molecular gas increases at lower metallicity as the CO molecule is more easily dissociated due to a lack of shielding dust \citep[e.g.,][]{Wolfire2010,Bolatto2013}.}

\change{Applying these previous results to the starburst-driven outflow of M\,82, we would expect a low CO-dark molecular gas fraction. The metallicity in this region is solar \citep[or slightly supersolar; e.g.,][]{Lopez2020}. At the resolution of these upGREAT observations, $A_V$ will vary substantially within a beam. Over their entire field-of-view, which mainly covers the central starburst, \citet{ForsterSchreiber2001} found $A_V=36\pm16$~mag. Extending the models from \citet{Wolfire2010} would imply a CO-dark gas fraction of $\sim10-25$\%. Because the shielding by dust in the outflow is likely lower than in the nucleus, we estimate that $\sim25$\% of the molecular gas may be in a CO-dark phase. Combining this with the fraction of the \CII\ line attributed to the CO-emitting molecular gas ($\sim20$\%), we would expect a $\sim5$\% contribution to the total \CII\ line from CO-dark molecular gas.}

\subsection{Ionized Gas in the Outflow}

The \CII\ line can also be excited in the ionized phase of the ISM. It is thought that, when \CII\ is excited in ionized conditions, the majority of the \CII\ emission arises from regions of diffuse ionized gas rather than \HII\ regions \citep[e.g.,][]{Nagao2011,Contursi2013}. In the KINGFISH sample, \citet{Croxall2017} found that $25\pm8$\% of the \CII\ is associated with the ionized gas. In star-forming regions in the center of the Milky Way, \citet{Harris2021} \change{found} that PDRs and \HII\ regions contribute roughly equally to the \CII\ flux. Therefore, while it is likely that the fraction of \CII\ associated with the ionized gas changes with environment, the ionized ISM is not the dominant contributor to the \CII\ emission.

\change{\citet{Contursi2013} found that the ionized gas traced by \Ha\ in the outflow of M\,82 is kinematically decoupled from the neutral (atomic and molecular) phases and from ionized gas traced by \OIII\ 88\micron\ emission. Moreover, they found that the ionized and neutral phases may not be co-spatial in the outflow. They proposed a scenario where the \Ha-emitting ionized gas is more extended, is confined to the walls of the biconical outflow, has a higher outflow velocity ($\sim$600~\kms), and is ionized (at least in part) by shocks between the outflowing X-ray-emitting plasma and the galaxy halo. The \OIII-emitting ionized gas, on the other hand, is more collimated (even narrower than the neutral material), has a slower outflow velocity ($\sim$75~\kms), and is primarily photoionized by the starburst.}

\change{For the most part, the \CII\ lines we measure have similar spectral shapes to the \HI\ and/or CO (e.g., Figures \ref{fig:columndensity_H0} and \ref{fig:columndensity_H2}), indicating that the \CII-emitting gas is likely coupled to the neutral material and photoionized by the starburst. We do not see evidence for components of the \CII\ line that do not correspond to either the \HI\ or CO, except perhaps the most redshifted edge of the \CII\ profile in Pixel 1 (Figures \ref{fig:CII_spectra}, \ref{fig:columndensity_H0}, and \ref{fig:columndensity_H0}). 
}

\change{Another method to determine the fraction of \CII\ associated with the ionized gas is to compare the \CII\ intensity to that of the \NII\ 205\micron\ line. The \NII\ 205\micron\ line is only excited in the ionized ISM and it has a similar critical density as the \CII\ 158\micron\ line. Using this method, \citet{Tarantino2021} found that the fraction of \CII\ arising from the ionized medium is $<12$\% in the disks of two normally star-forming galaxies.} Unfortunately, there are no observations of the \NII\ 205\micron\ line from either \herschel\ or SOFIA in M\,82. \citet{Contursi2013} used the \NII\ 122\micron\ line to place limits on the fraction of \CII\ arising from the ionized medium in the outflow at distances $<1$~kpc from the disk. Because the \CII\ and \NII\ 122\micron\ lines do not have similar critical densities, these results are dependent on the electron density, which is largely unknown in the outflow of M\,82 \citep[e.g.,][]{Shopbell1998,Yoshida2011}. Nevertheless, \citet{Contursi2013} were able to constrain that $\sim11-40$\% of \CII\ emission arises from the ionized gas in the southern outflow of M\,82 (at distances $<1$~kpc from the midplane). They note that these estimates will be even more uncertain when applied to regions with a complex mix of different ISM components, which is certainly the case in the outflow of M\,82. We can conclude, however, that the ionized gas is not the dominant contributor to the \CII\ emission in the outflow of M\,82.

\begin{figure*}
    \centering
    \includegraphics[width=\textwidth]{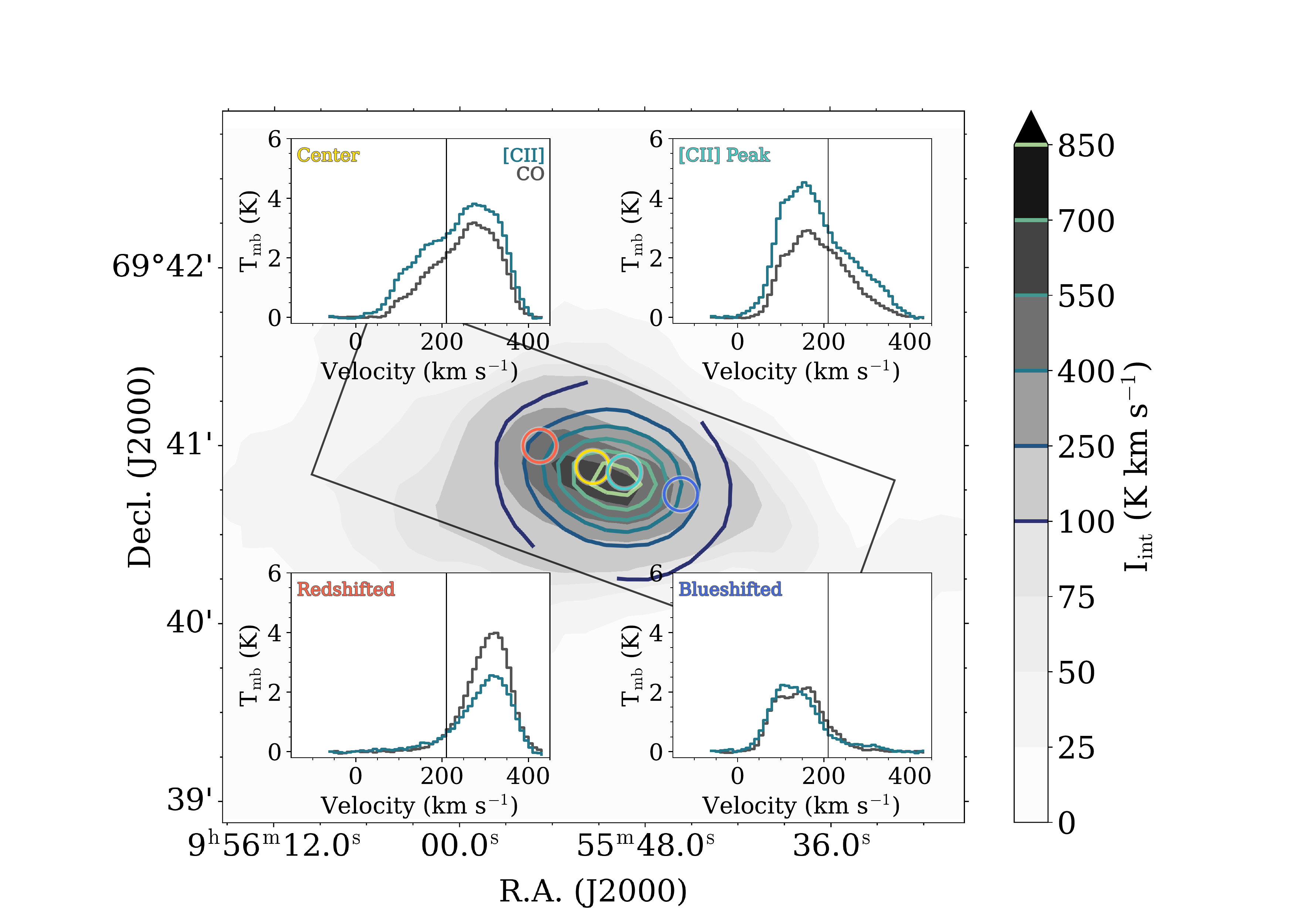}
    \caption{A comparison of the \CII\ and CO spectra in the disk of M\,82. The grayscale filled contours show the CO integrated intensity. The blue-green open contours show the \CII\ integrated intensities (see the lines in the color bar). The black rectangle marks the fully-sampled region of the \CII\ map. The subplots show the spectra of the CO (gray) and \CII\ (teal) extracted at the galaxy center (yellow), at the peak of the \CII\ emission (cyan), and on the redshifted and blueshifted sides of the galaxy (red, blue). Both the CO and \CII\ intensities are reported as main-beam brightness temperatures, and the data sets have been matched to the same spatial and spectral gridding and convolved to the CO beam size (22\arcsec). The black vertical lines in the panels show the \vsys$~=~$210~\kms. Comparisons with \HI\ are not possible to due strong absorption over this region (see Section \ref{sssec:HI}).}
    \label{fig:disk_spectra}
\end{figure*}

\subsection{\change{Synthesis}}
\change{We summarize the contributions to the \CII\ emission in the outflow of M\,82 from various ISM phases as follows. From our comparison of the \htwo\ column density measured from CO compared to that predicted from \CII, we find that $\sim20$\% of the \CII\ emission may arise from CO-emitting molecular gas (Figure \ref{fig:columndensity_H2}). We attempt to quantify the contribution of CO-dark molecular gas, finding $\sim4$\% of the total \CII\ emission may arise from this phase. Therefore, $\sim24$\% of the \CII\ emission in the outflow of M\,82 may arise from the molecular ISM. For the contribution of ionized gas, we rely on measurements by \citet{Contursi2013} and estimate that $\sim20$\% of the total \CII\ emission may arise from the ionized ISM. Finally, we compare the \HI\ column density measured from \HI\ data to that predicted from \CII\ and find that there is sufficient \HI\ to fully explain the \CII\ emission (Figure \ref{fig:columndensity_H0}). Therefore, we can attribute the remaining $\sim55$\% of the \CII\ emission to the atomic ISM.}

\section{\CIIsect\ in the Disk of M\,82}
\label{ssec:results_disk}

As shown in Figures \ref{fig:composite_map} and \ref{fig:CII_maps}, we detect the \CII\ line at high significance in the disk of M\,82. We calculate a total \CII\ mass in the disk of $4\times10^6$~\msun, calculated where SNR~$>2$ (Figure \ref{fig:CII_maps}) following Equations \ref{eq:crawford_simp} and \ref{eq:MCp}. 

In Figure \ref{fig:disk_spectra}, we compare representative \CII\ and CO spectra in the disk of M\,82 at the center, at the peak of the \CII\ emission, and on the redshifted and blueshifted sides of the galaxy. For this figure, we have spectrally smoothed the CO data to 10~\kms\ to match the \CII, convolved the \CII\ data to the larger 22\arcsec\ CO beam, and matched the pixel sizes. Both the CO and \CII\ are reported in main-beam brightness temperature units. We calculate the integrated intensity (moment 0) of the \CII\ and CO on these matched data sets, as shown in Figure \ref{fig:disk_spectra}. Comparisons with the \HI\ spectra are not possible because of deep absorption features in the \HI\ over the region where the \CII\ is detected in the disk \citep[Section \ref{sssec:HI}, Figure \ref{fig:composite_map}, and][]{Martini2018}. In general, we find that the CO and \CII\ intensities and spectral shapes are quite similar, in agreement with previous work in this galaxy at lower resolution \citep[e.g.,][]{Stacey1991}.

\begin{figure}
    \centering
    \includegraphics[width=\columnwidth]{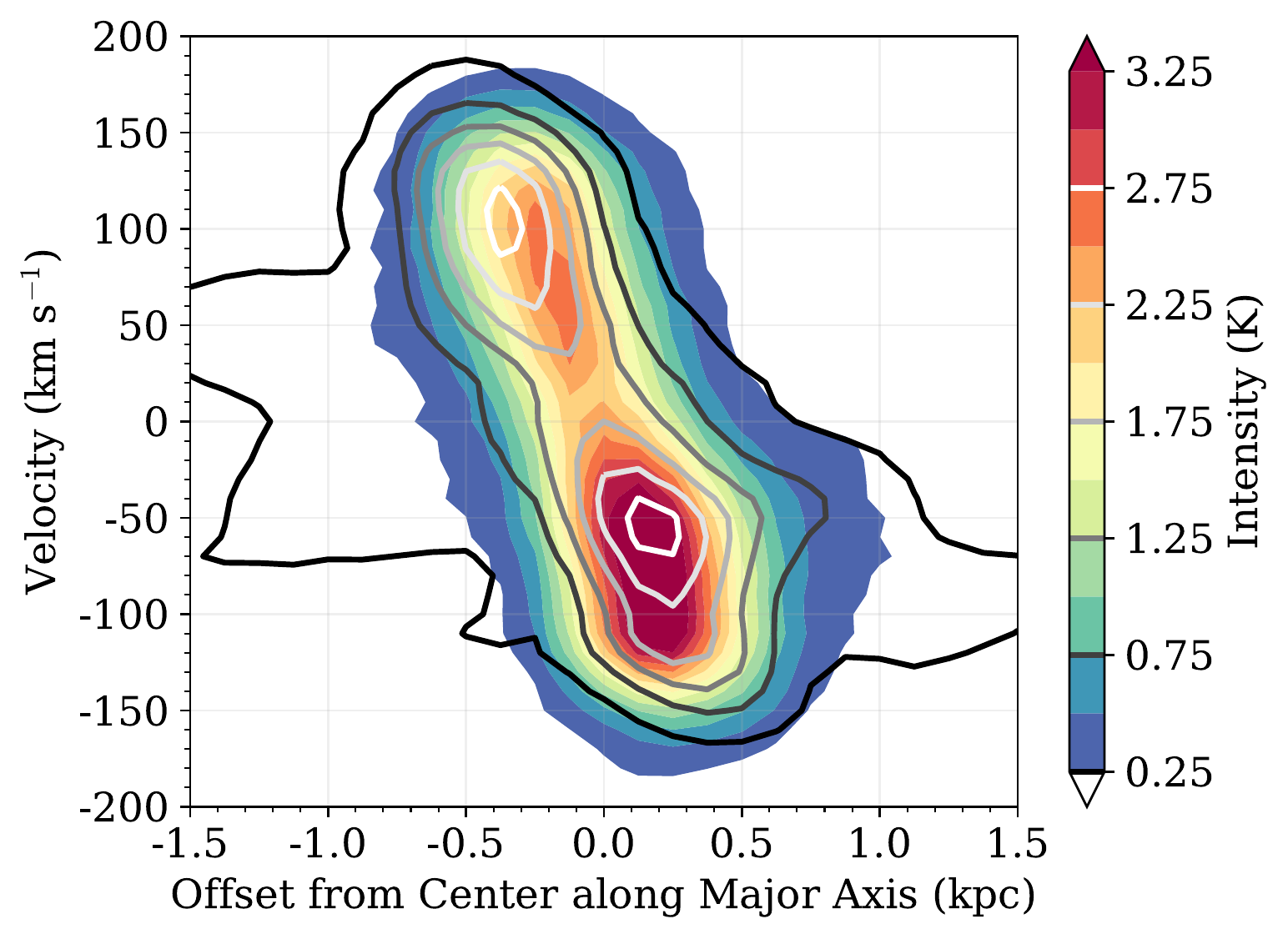}
    \caption{A PV diagram of the \CII\ (filled colored contours) and CO (open grayscale contours) in the center of M\,82. This slice is extracted along the galaxy major axis and covers the entire fully-sampled region of the \CII\ map ($3\times1$~kpc). The CO and \CII\ data have the same beam size, pixel scale, and velocity resolution. The $x$-axis shows the offset along the galaxy major axis with respect to the center. The $y$-axis shows the velocity relative to \vsys~$=210$~\kms\ \citep{Krieger2021}. In general, the kinematics of the CO and \CII\ agree, though the \CII\ velocity rises faster than the CO in the center. }
    \label{fig:PV}
\end{figure}

In Figure \ref{fig:PV}, we make a position-velocity (PV) diagram of the matched-resolution CO and \CII\ in the disk of M\,82. The velocities for both datasets are reported in the radio velocity convention and in the LSRK frame. We extract the \CII\ and CO PV slices over the entire fully-sampled region of the \CII\ map, covering the central $3{\rm\ kpc\ }\times1{\rm\ kpc}$ of the galaxy along its major axis (Section \ref{ssec:diskmap}). We collapse the PV slice along the minor axis by taking an average weighted by the intensity of each spaxel. Spaxels with intensities less than the rms noise of the cube (in areas away from emission) are removed.

As shown in Figure \ref{fig:PV}, the kinematics of the \CII\ and CO generally agree in the inner regions of M\,82. In detail, however, there are some differences. The \CII\ velocity appears to rise faster in the center compared to the CO. This disagreement is worse on the redshifted (i.e. eastern) side of the galaxy. The CO on this side of the galaxy appears more kinematically disturbed (e.g., Figure 2 of \citealt{Leroy2015} and Figure 2 of \citealt{Krieger2021}). The \HI\ on the other hand appears {\em less} kinematically disturbed on this side of the galaxy (e.g., Figure 1 of \citealt{Martini2018}). Therefore, if the \CII\ in the disk primarily arises from the atomic gas (as it does in the outflow) then perhaps this could explain the kinematic differences we see in the PV diagrams, though this is somewhat speculative. 

\section{\CIIsect\ and CO throughout M\,82}
\label{sec:M82_disk_outflow}

\subsection{\CIIsect\ and CO Luminosity Ratios}

\begin{figure}
    \centering
    \includegraphics[width=\columnwidth]{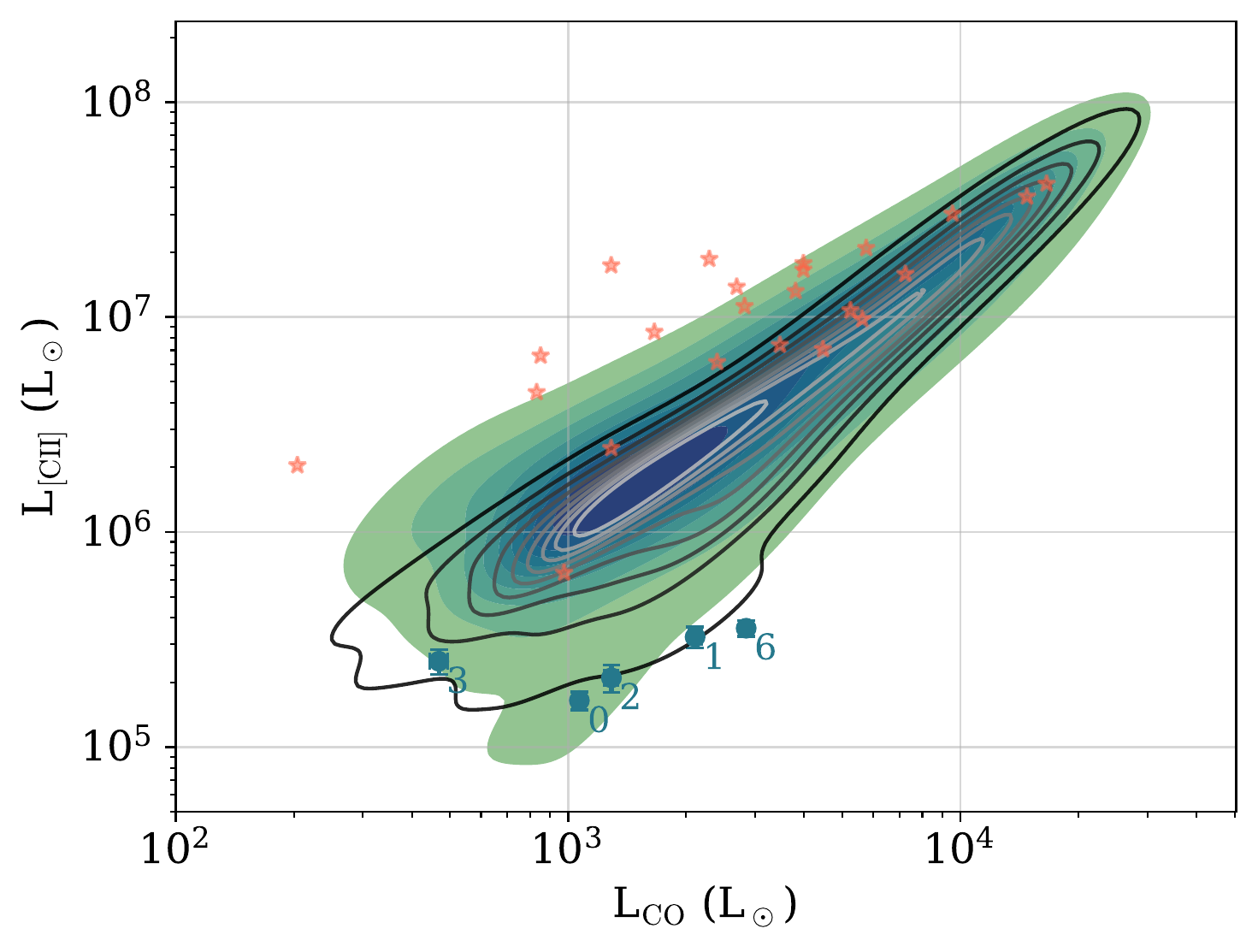}
    \caption{
    \change{The CO and upGREAT \CII\ luminosities in the disk (green-blue filled contours) and outflow (teal circles) of M\,82. The teal numbers indicate the pixel number of each outflow pointing. We compare our measurements to those from PACS \citep{Contursi2013} in the grayscale contours. For all contours, the colors reflect the density of points. We also compare to integrated measurements from the xCOLD GASS survey of star-forming galaxies \citep[red stars;][]{Accurso2017b}.}
    }
    \label{fig:LCII_LCO}
\end{figure}

A somewhat different angle to assess the contribution of the molecular gas to the \CII\ emission than presented in Section \ref{ssec:molgas_outflow} is to compare the luminosity ratios of \CII\ and CO. In principle, this ratio is sensitive to the FUV radiation field and the ability of CO to self-shield via dust from the FUV radiation \citep[e.g.,][]{Accurso2017b}. To calculate the luminosities from the integrated intensities, we follow Equation 1 of \citet{Solomon1997}:
\begin{equation}
    \label{eq:L_from_intinten}
    L = 1.04\times10^{-3}\left[\frac{\rm I_{int}}{\rm Jy~km~s^{-1}}\right]\left[\frac{\nu(1+z)^{-1}}{\rm GHz}\right]\left[\frac{D_L}{\rm Mpc}\right]^2~L_\odot
\end{equation}
where $\nu$ is the line rest frequency, $z$ is the redshift, and $D_L$ is the luminosity distance. Because M\,82 is very nearby, we take $D_L=3.63$~Mpc and calculate $z = V_{\rm sys}/c$ with \vsys~$=210$~\kms\ as the recessional velocity. We show these luminosities in Figure \ref{fig:LCII_LCO} for both the disk (blue-green filled contours) and outflow (teal circles; Table \ref{tab:outflowfits_HICO}) of M\,82.

\change{We compare the \LCII\ we infer from our measurements to those measured by \citet{Contursi2013}. The region mapped by PACS extends farther into the outflow ($\sim1$~kpc) compared to the upGREAT map ($\sim0.5$~kpc), but not as far as the upGREAT outflow pointings. We match the PACS \CII\ data to the IRAM 30-m CO data by first convolving the PACS cube to a 22\arcsec\ Gaussian beam using the kernel provided by \citet{Aniano2011}. We then match the pixel scales of the \CII\ and CO cubes and re-derive the integrated intensities. We show the PACS \CII\ and matched CO luminosities in Figure \ref{fig:LCII_LCO} as the grayscale contours. The PACS and upGREAT data agree well (as expected from the analysis in Section \ref{ssec:diskmap}).}

We compare the ratios we derive for M\,82 to a sample of 24 normally star-forming galaxies from the xCOLD GASS survey from \citet{Accurso2017b} in Figure \ref{fig:LCII_LCO}. In the disk of M\,82 we find somewhat lower \LCII/\LCO\ ratios than for normally star-forming galaxies (the median for M\,82 is 0.4~dex \change{lower}). We note, however, that the measurements from \citet{Accurso2017b} are galaxy-integrated, whereas the measurements from the disk of M\,82 are in ${\rm\approx(125~pc)^2}$ regions (one pixel). 

Another important caveat is that the galaxies studied by \citet{Accurso2017b} tend to have metallicities less than solar, whereas M\,82 has solar (or slightly supersolar) metallicity \citep[e.g.,][]{Lopez2020}. \citet{Accurso2017b} found the strongest trend in \LCII/\LCO\ with metallicity (in 12+log(O/H) units), where higher metallicity systems have lower \LCII/\LCO. They also found a strong trend with the hardness of the radiation field (defined as the ratio of the FUV to near UV (NUV) flux), where harder radiation environments have lower \LCII/\LCO. Both of these trends may help explain why the disk of M\,82 has lower \LCII/\LCO\ than the star-forming galaxies analyzed by \citet{Accurso2017b}.

\subsection{Fraction of \CIIsect\ in the Outflow of M\,82}
\label{ssec:cii_frac_outflow}

At ${z\sim3-7}$, recent observations with ALMA \change{have revealed} extended \CII\ halos around some galaxies extending up to ${\sim10}$~kpc from the center and containing $\sim50$\% of the \CII\ emission \citep[e.g.,][though see also \citealt{Novak2020}  for counter-examples]{Rybak2019,Fujimoto2019,Fujimoto2020,Ginolfi2020,Meyer2022}. We note that these studies are a mix of detections from individual systems and stacks, as well as spatially unresolved and marginally-resolved studies. Simulations have shown that these extended \CII\ halos can be powered by supernova-driven cooling outflows \citep[e.g.,][]{Pizzati2020}.

As M\,82 is sometimes used as a local anchor for high-z star-forming galaxies, it is interesting to constrain the fraction of \CII\ emission in the outflow compared to the disk. The CO in the outflow of M\,82 only extends for $\sim2-3$~kpc above and below the midplane \citep{Walter2002,Salak2013,Leroy2015,Krieger2021}. \citet{Martini2018} found that the \HI\ is significantly more extended, reaching $\sim5$~kpc above the midplane and $\sim10$~kpc below (in the direction of M\,81 with which M\,82 is interacting; see also \citealt{Yun1994}). Based on lower resolution CO data, \citet{Walter2002} measured $3.3\times10^8$~\msun\ of molecular gas in the halo and outflow of M\,82, $2.3\times10^8$~\msun\ in the disk, and $8.0\times10^8$~\msun\ in the tidal streamers, for a total molecular gas mass of $1.3\times10^9$~\msun. We note that this total molecular gas mass agrees with other more recent measurements \citep{Salak2013,Leroy2015,Krieger2021}. Overall, \citet{Walter2002} found that while $>70$\% of the molecular material resides outside of the disk of M\,82, only $\sim25$\% of the total molecular gas mass is swept up in the outflow/halo component with the rest being in the tidal streamers.

For the measurements of the \CII\ halos at high-z, outflow/halo and streamer components would be mixed together. However, as we know from M\,82, not all of this mass is outflowing so attributing all of the extended \CII\ emission to the outflow can significantly overestimate the \CII\ mass outflow rates in these high-z systems. This rough comparison assumes that the \CII\ and CO masses in each component track one another.

Because our upGREAT observations do not cover the full extent of the outflow of M\,82, we cannot directly measure the total \CII\ extent, mass, or flux in the outflow relative to the disk. We will instead extrapolate our \CII\ measurements to infer the total fraction of \CII\ we might expect based on the CO. In the outflow, the average ratio of the peak brightness of the CO and \CII\ line (where \CII\ is detected) is $2.0\pm0.8$, where the uncertainty is the standard deviation (see Figure \ref{fig:CII_spectra}). In the central disk, the peak brightness ratio is nearly the same, with an average and standard deviation of $2.1\pm1.1$ (see e.g., Figure \ref{fig:disk_spectra}). We note that \citet{Walter2002} define the M\,82 disk as the inner 1~kpc, which is very similar to the region of the disk where we robustly detect \CII\ emission (e.g., Figure \ref{fig:CII_maps}). Therefore, since the ratio of the intensities in the central disk and outflow are roughly the same, we might also expect the relative mass ratios to be the same as well. This means that we would expect to find $\sim25$\% of the total \CII\ in the outflow, $\sim18$\% in the inner disk, with the remaining \CII\ distributed in the streamers. Given that we measure a \CII\ mass of $4\times10^6$~\msun\ in the disk (Section \ref{ssec:results_disk}), we would predict $\sim5.5\times10^6$~\msun\ of \CII\ in the entire outflow of M\,82. This mass corresponds to a total integrated intensity of $\sim3.5\times10^4$~\Kkms\ (Equations \ref{eq:MCp} and \ref{eq:crawford_simp}) and \LCII~$\sim1.7\times10^8$~\lsun\ (Equation \ref{eq:L_from_intinten}). 

Indeed, some observational studies of \CII\ halos at high-z do find evidence of an extended component but without a broad \CII\ profile that would indicate an outflow \citep[e.g.,][]{Novak2020,Spilker2020,Meyer2022}. In particular, \citet{Spilker2020} studied molecular outflows in a sample of lensed dusty star-forming galaxies at $z>4$. They found that $>70$\% of the galaxies in their sample had clear evidence for a molecular outflow based on OH 119\micron\ absorption. However, none of these galaxies with confirmed molecular outflows had broad \CII\ emission line wings. This suggests that, at least in this population of $z>4$ highly star-forming galaxies, that \CII\ is not a robust tracer of outflowing molecular gas \change{at high redshift}. %From our comparison with M\,82, it is possible that while there may be \CII\ in these outflows, the fraction by mass is small and hence difficult to measure.

\change{In summary, in M\,82 we clearly detect \CII\ in the starburst-driven outflow, though we expect that the outflowing \CII\ accounts for only $\sim25$\% of the total \CII\ of the system. This is somewhat different than is observed for high redshift \CII\ outflows, where there is evidence for molecular outflows in broad emission lines but that lack robust \CII\ \citep[e.g.,][]{Novak2020,Spilker2020,Meyer2022}.}

\section{FUV Radiation Field}
\label{ssec:radiationfield}

\begin{figure*}
    \centering
    \includegraphics[width=0.535\textwidth]{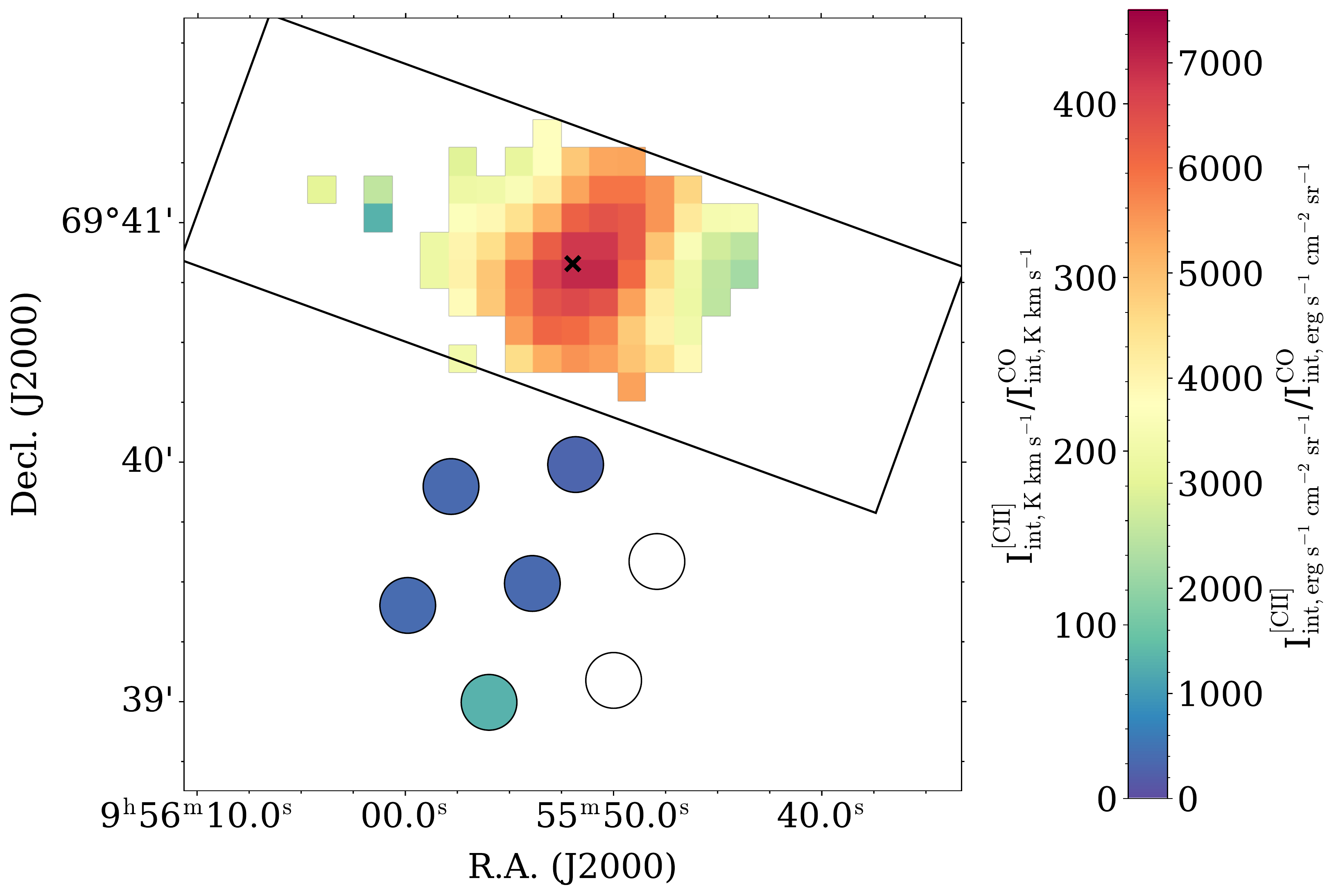}
    \includegraphics[width=0.455\textwidth]{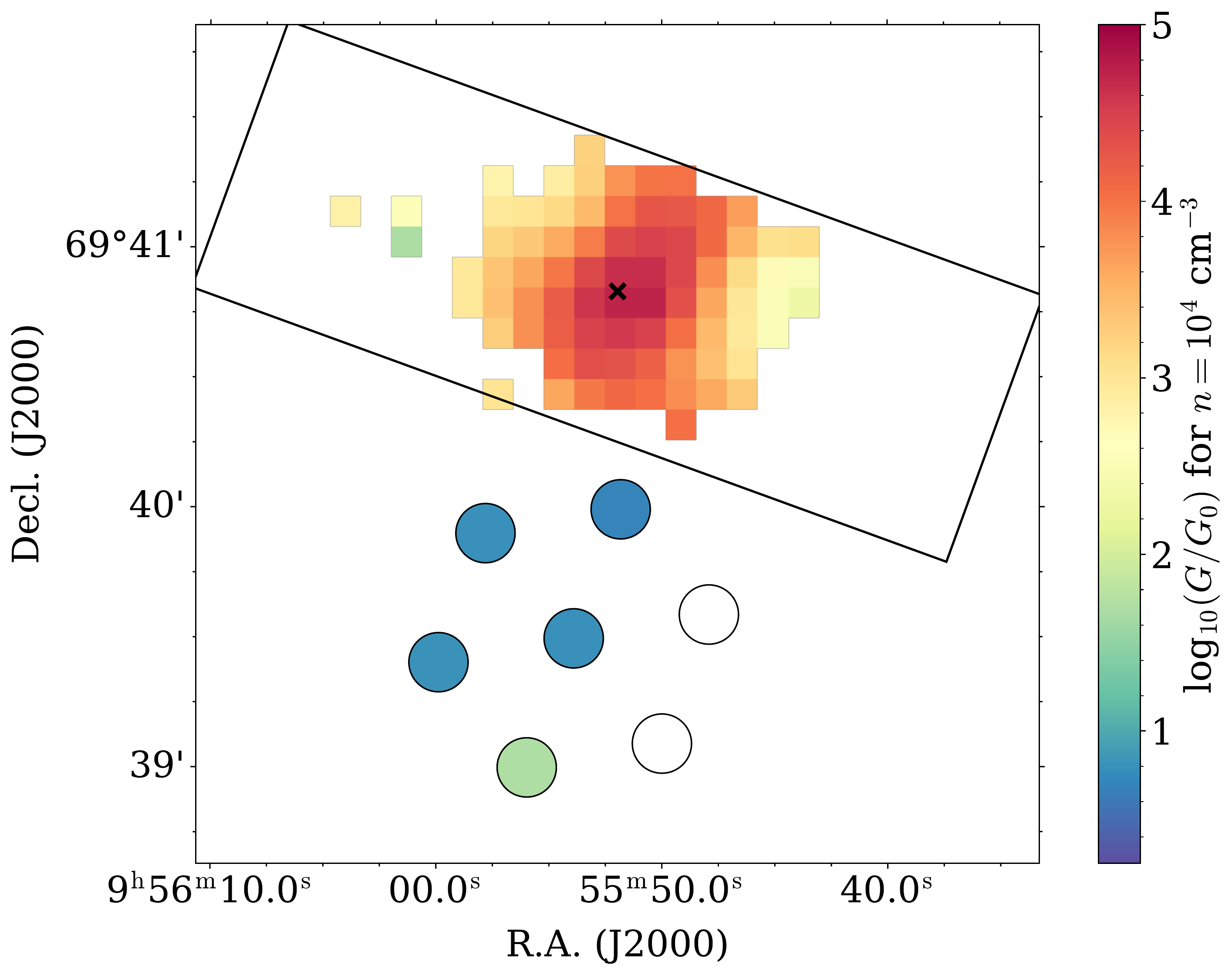}
    \caption{Left: The ratio of the \CII\ to CO integrated intensities (in units of \ergscmsr\ and \Kkms), in the same style as Figure \ref{fig:CII_maps}. In the disk, the \CII\ and CO data cubes, from which the integrated intensities are calculated, have the same beam size, pixel scale, and velocity resolution (see Section \ref{ssec:results_disk}). Right: The estimated FUV radiation field strength based on the PDR model by \citet{Kaufman1999} using the ratios from the left plot and assuming a density of $10^4$~cm\pert.}
    \label{fig:CII_CO_ratio}
\end{figure*}

Within PDRs, photoelectric heating of small dust grains efficiently heats the region and governs the chemistry, and this heating is primarily governed by the density ($n$) and the far ultraviolet (FUV) radiation field strength \citep[$G/G_0$\footnote{$G_0$ is the Habing field for radiation with energies from $6-13.6$~eV, equivalent to $1.6\times10^{-3}$~erg~s\per~cm\pers.}; e.g.,][]{Tielens1985,Wolfire1990}. The intensities of lines emitted in a PDR are, therefore, sensitive to these properties as well, and line ratios of FIR fine structure lines and CO can be used to constrain $n$ and $G$ \citep[e.g.,][]{Wolfire1990,Kaufman1999}. In the center of M\,82, \citet{Kaufman1999} applied their PDR model to integrated measurements of FIR lines, finding $n\sim10^4$~cm\pert\ and $G\sim10^{3.5}~G_0$. \citet{Contursi2013} also used the \citet{Kaufman1999} PDR model to measure the FUV radiation field in the central starburst and southern outflow of M\,82 (within 1~kpc of the midplane), finding $G=10^{3.1-3.4}~G_0$ in the starburst and $G\sim10^{2.1}~G_0$ in the outflow.

\subsection{Radiation Field Constraints from CO and \CIIsect}

From their PDR model, \citet{Kaufman1999} find that, while the ratio of the \CII\ to CO integrated intensities (where both quantities are in units of \ergscmsr) is mostly sensitive to the column density of \Cp\ and the temperature, this ratio does also depend on $n$ and $G$ (see their Figure 9). We, therefore, use the line ratios of \CII\ and CO that we measure in the disk and outflow of M\,82 to place new constraints on the FUV radiation field in this region.

Figure \ref{fig:CII_CO_ratio} (left) shows the integrated intensity ratio of \CII\ to CO. Like has been found previously \citep[e.g.,][]{Stacey1991,Kaufman1999}, ratios range from $\sim~3-7\times10^3$ in the disk of M\,82. Ratios in the outflow are substantially lower, with Pixels 0, 1, 2, and 6 (purple colors in Figure \ref{fig:CII_CO_ratio} left) having ratios of $300-400$.

Using the PDR model developed by \citet{Kaufman1999}, we place limits on the strength of the FUV radiation field assuming some density of the material. We interpolate the predictions of the \citet{Kaufman1999} PDR model (their Figure 9) to estimate $G$ for a given \CII-to-CO intensity ratio. This estimation of $G$ is shown in Figure \ref{fig:CII_CO_ratio} (right). For a fiducial density of $n=10^4$~cm\pert\ (following the results of \citealt{Kaufman1999}), the observed \CII/CO ratios in the disk can be explained by $G\sim10^{3-5}~G_0$, in agreement (though somewhat circularly) with \citet{Kaufman1999} and \citet{Contursi2013}.

In the outflow, we estimate a much lower FUV radiation field, with Pixels 0, 1, 2, and 6 (blue colors in Figure \ref{fig:CII_CO_ratio} right) having $G\sim6~G_0$. Although the outflow is less dense than the disk \citep[e.g.,][]{Yuan2023}, the result of a much smaller FUV radiation field holds (see Section \ref{ssec:density_uncert} for more discussion on the effect of uncertainties in the assumed density). 

\subsection{Uncertainties on $G/G_0$ Due to the Assumed Density}
\label{ssec:density_uncert}

A major source of uncertainty in these calculations is the assumed density ($n$). While we assume a fiducial $n=10^4$~cm\pert\ motivated by the results of \citet{Kaufman1999} in the center of M\,82, it is unlikely that the outflow is this dense (as discussed in Section \ref{ssec:results_outflow} and in \citealt{Contursi2013}). 

In the disk, we allow the assumed density to vary by 0.5~dex (i.e., $n=10^{3.5-4.5}$~cm\pert). Because the PDR models are not monotonic with density (see Figure 9 of \citealt{Kaufman1999}), we perform a grid search in steps of 0.1~dex in density and find the minimum and maximum values of $G/G_0$ at each disk pixel. For the disk, the uncertainty at each pixel is roughly the same ($\log_{10}(G/G_0)^{+0.5}_{-0.6}$). 

In the outflow, the density is almost certainly much lower than $10^4$~cm\pert. We employ the same grid search as described above over a range of $n=10^{2-4}$~cm\pert. The lower limit encompasses the density limits determined by \citet{Contursi2013} and used in Section \ref{ssec:atomgas_outflow}. We assume that our fiducial density is the maximum density in the outflow. The lower uncertainty on $G/G_0$ in the outflow comes from allowing the assumed density to be 2~dex lower than the fiducial assumption (i.e., $n=10^{2-4}$~cm\pert). 

Accounting for the density uncertainties, we find $G/G_0\approx10^{1.8-5.2}$ in the disk and $G/G_0\approx10^{0.1-1.0}$ in the outflow (excluding each green point in the disk and outflow). Therefore, even considering the uncertainties from the density assumptions, the radiation field in the outflow is substantially lower than in the disk.

\section{Summary}
\label{sec:summary}

M\,82 is an archetypal example of a starburst-driven outflow and is an ideal laboratory to study the detailed physics of superwinds. Here, we present new velocity-resolved observations of the \CII\ 158\micron\ emission line towards the center and southern outflow of M\,82, enabled by upGREAT onboard SOFIA. With upGREAT, we mapped the central 3~kpc~$\times$~1~kpc of the disk of M\,82. In the southern outflow, we use one pointing of the seven-pixel upGREAT array to measure \CII\ at distances of $1-2$~kpc from the midplane (Figure \ref{fig:composite_map}). Below we summarize the main results of this analysis, indicating the relevant figures and/or tables:

\begin{enumerate}
    \itemsep0em
    \item \change{We detect the \CII\ line out to 2~kpc from the midplane along the southern outflow at 10~\kms\ velocity resolution (Figure \ref{fig:CII_spectra}). This twice as far as previously probed by \herschel\ PACS \citep{Contursi2013,Herrera-Camus2018}.}
    \item We compare the column densities of the atomic medium measured from the \HI\ data to the CNM column density measured from the \CII\ spectra (Figure \ref{fig:columndensity_H0}). \change{Similarly, we compare the column densities of the molecular medium measured from the CO data to the \htwo\ column density measured from the \CII\ spectra (Figure \ref{fig:columndensity_H2}). From these comparisons, we find that the majority ($>55$\%) of the \CII\ arises from the atomic component. It is likely that the molecular gas (including an estimate of the CO-dark molecular gas) contributes $\sim25$\% and that the ionized gas contributes $\sim20$\% of the \CII\ emission.} 
    \item \change{We are able to extend the results from \citet{Walter2002} from CO to estimate the total fraction of \CII\ in the outflow of M\,82. While the bulk of the \CII\ emitting gas is likely outside of the main disk, only a small fraction is actually outflowing (with the rest located in tidal streamers, for example). This may help inform observations of \CII\ halos at higher redshifts, which sometimes lack outflow signatures.}
    \item We estimate the strength of the FUV radiation field in the disk and outflow of M\,82 using the PDR model developed by \citet{Kaufman1999}. In the disk of M\,82, we find $G\sim10^{3-5}~G_0$, in agreement with previous measurements \citep[Figure \ref{fig:CII_CO_ratio};][]{Kaufman1999,Contursi2013}. The FUV radiation field we measure $\sim1.5$~kpc away from the disk in the outflow, however, is 2-3 orders of magnitude lower than in the disk. 
\end{enumerate}

Owing to the sensitivity and wavelength coverage of ALMA, the \CII\ 158\micron\ emission line is routinely observed in galaxies at $z\gtrsim2$. Because this line is bright, it is a more attractive tracer of molecular gas than CO in these systems. However, it is crucial to understand the contribution of the various ISM phases to the \CII\ line in order to use this line as a tracer of molecular gas and star formation. The galaxy systems studied so far at $z\gtrsim2$ tend to have high star formation rates, so understanding the behavior of the \CII\ line in this starburst environment is critical to inform these high-$z$ measurements. Unfortunately, with the end of the SOFIA mission, observations of \change{\CII\ and other FIR lines} in the local Universe will be possible only with balloon missions for at least the next few decades. Future  facilities in space are needed to more completely understand how the various ISM phases contribute to the \CII\ 158\micron\ line as a function of \change{spatial resolution, environment, and ISM conditions.}

\begin{acknowledgments}
R.C.L. acknowledges support for this work provided by NASA through award number 08-0225 issued by the Universities Space Research Association, Inc. (USRA) and by a National Science Foundation (NSF) Astronomy and Astrophysics Postdoctoral Fellowship under award AST-2102625. \change{A.D.B. acknowledges support from the NSF under award AST-2108140.} \changes{R.H.-C. thanks the Max Planck Society for support under the Partner Group project "The Baryon Cycle in Galaxies" between the Max Planck for Extraterrestrial Physics and the Universidad de Concepción. R.H-C. also gratefully acknowledge financial support from Millenium Nucleus NCN19058 (TITANs), and ANID BASAL projects ACE210002 and FB210003.}
This work is based on observations made with the NASA/DLR Stratospheric Observatory for Infrared Astronomy (SOFIA), project \#08\_0225. SOFIA is jointly operated by USRA, under NASA contract NNA17BF53C, and the Deutsches SOFIA Institut (DSI) under DLR contract 50 OK 2002 to the University of Stuttgart.
This work is based on observations carried out under project No. 107-19 with IRAM 30 m telescope. IRAM is supported by INSU/CNRS (France), MPG (Germany), and IGN (Spain).
The National Radio Astronomy Observatory is a facility of the National Science Foundation operated under cooperative agreement by Associated Universities, Inc.
This research has made use of the NASA/IPAC Extragalactic Database (NED), which is funded by the National Aeronautics and Space Administration and operated by the California Institute of Technology.
This research has made use of NASA's Astrophysics Data System Bibliographic Services.
\end{acknowledgments}

\facilities{SOFIA (upGREAT), IRAM:30m, VLA\change{, GBT}}

\software{Astropy \citep{Astropy2018,Astropy2022}, {\em kalibrate} \citep{Guan2012}, MatPlotLib \citep{matplotlib}, NumPy \citep{numpy}, pandas \citep{pandas}, photutils \citep{photutils}, SciPy \citep{scipy}, seaborn \citep{seaborn}, spectral-cube \citep{spectralcube}, WebPlotDigitizer \citep{WebPlotDigitizer}}

\bibliographystyle{aasjournal}

\appendix

\section{Calculating the C$^+$ Density and Mass in the Outflow}
\label{app:density_calc}

We calculate the \CII\ density and mass in the outflow of M82 channel-by-channel for the velocity-resolved \CII\ spectrum. We describe this calculation below and direct the reader to \citet{Goldsmith2012} and \citet[][and references therein]{Tarantino2021} for a much more complete discussion. We note that these calculations assume the \CII\ is optically thin, which is well supported by the results of \citet{Contursi2013}.

First, we can relate the column density of \Cp\ (\NCp) to the \CII\ intensity ($I_{\rm[CII]}$) in each channel of the spectrum:

% \begin{equation}
% \begin{split}
%     \label{eq:crawford}
%     I_{\rm [CII]}\Delta V = 2.3\times10^{-21}B\left[\frac{N_{\rm C^{+}}}{\rm cm^{-2}}\right] {\rm\ erg\ s^{-1}\ cm^{-2}\ sr^{-1}}\\
%     = 3.3\times10^{-16} B \left[\frac{N_{\rm C^{+}}}{\rm cm^{-2}}\right] {\rm K\ km\ s^{-1}}
% \end{split}
% \end{equation}
\begin{equation}
\begin{split}
    \label{eq:crawford}
    N_{\rm C^{+}} = \frac{3.0\times10^{15}}{B}\left[\frac{I_{\rm [CII]}}{\rm K}\right] \left[\frac{\Delta V}{\rm km~s^{-1}}\right] {\rm\ cm^{-2}}
\end{split}
\end{equation}
where
\begin{equation}
    \label{eq:B}
    B = \frac{2e^{-91.2/T}}{1+2e^{-91.2/T}+A_{ul}/(\Sigma R_{ul}n_i)}
\end{equation}
and where $\Delta V$ is the channel width in \kms, $T$ is the kinetic temperature in K, $A_{ul}$ is the Einstein A spontaneous decay rate ($2.3\times10^{-6}$~s\per\ for the 158~\micron\ transitions of \CII), and $\Sigma R_{ul}n_i$ is the sum over all $i$ collisional partners with collisional decay rates $R_{ul}$ and volume densities $n_i$ \citep{Crawford1985,Goldsmith2012,Tarantino2021}.

\subsection{Collisions with Atomic Gas}
\label{app:atomic}
First, we focus on collisions with the atomic gas only. In particular, the \CII\ is excited primarily in the cold neutral medium \citep[CNM; e.g.,][]{Pineda2013,Fahrion2017,Herrera-Camus2017,Tarantino2021}. Therefore, the sum over the collisional partners in Equation \ref{eq:B} can be simplified to include only neutral hydrogen and helium:
\begin{equation}
\begin{split}
    \label{eq:sum}
    \Sigma R_{ul,i}n_i = n_{\rm CNM}\left[R_{ul}({\rm H}^0)+R_{ul}({\rm He}^0)\right] {~\rm s^{-1}} \\ = 1.038R_{ul}({\rm H}^0)n_{\rm CNM} {~\rm s^{-1}}
\end{split}
\end{equation}
where we have made the final simplification because the collisional rate for helium is 38\% of that for hydrogen \citep{Draine2011}.
\citet{Goldsmith2012} calculated that
\begin{equation}
    \label{eq:Rul}
    R_{ul}({\rm H}^0) = 4.0\times10^{-11}\left(16+0.35T^{0.5}+48T^{-1}\right) {\rm cm^3~s^{-1}}.
\end{equation}
We assume $T=250$~K and $n = 100~{\rm cm^{-3}}$ following the results of \citet[][the discussion in Section \ref{ssec:results_outflow}]{Contursi2013}. Therefore, $R_{ul}({\rm H}^0) = 9.1\times10^{-10}~{\rm cm^3~s^{-1}}$, $\Sigma R_{ul,i}n_i = 9.4\times10^{-8}$~s\per, and $B = 5.7\times10^{-2}$. With these assumptions, Equation \ref{eq:crawford} becomes
\begin{equation}
    \label{eq:crawford_simp}
    N_{\rm C^{+};\,H^{0}} = 5.2\times10^{16}\left[\frac{I_{\rm [CII]}}{\rm K}\right] \left[\frac{\Delta V}{\rm km~s^{-1}}\right] {\rm\ cm^{-2}}
\end{equation}
considering only collisions with the atomic gas.

From \NCpHo, we can estimate the effective CNM column density (\NCNMCII) based on the relative abundance of carbon to hydrogen (${\rm C/H} = 1.5\times10^{-4}$; \citealt{Gerin2015}) assuming all of the carbon is singly-ionized and all the hydrogen is atomic:
\begin{equation}
    \label{eq:N_HI_CII}
    N_{\rm CNM}^{\rm [CII]} = \frac{N_{\rm C^{+};\,H^{0}}}{{\rm C/H}}.
\end{equation}
The warm phase of the \HI\ accounts for $30-70$\% of the \HI\ emission \citep{Heiles2003} but does not contribute to the \CII\ emission \citep[e.g.,][]{Pineda2013,Fahrion2017,Herrera-Camus2017}. \change{We show this effective CNM column density profile based on the \CII\ in Figure \ref{fig:columndensity_H0} (teal).} %We compare these to the measured \HI\ column density profiles from the VLA data (see Section \ref{ssec:atomgas_outflow}) in magenta. In the upper right corner of each panel, \change{we report the ratio of \NCNMHI$/$\NCNMCII\ at the peaks of each profile}, calculated where the \HI\ has positive signal (roughly $25-300$~\kms, shown in the shading). The ratios are from $2-4$ (excluding Pixels 4 and 5 where there is no significant \CII\ signal), meaning that in general there is enough \HI\ to fully explain the \CII.

\subsection{Collisions with Molecular Gas}
\label{app:mol}
Next, we consider collisions with the molecular gas, \htwo. In this case, the sum over the collisional partners in Equation \ref{eq:B} can be simplified to include only molecular hydrogen:
\begin{equation}
    \label{eq:sum_H2}
    \Sigma R_{ul,i}n_i = n_{\rm mol}R_{ul}({\rm H}_2) {~\rm s^{-1}}.
\end{equation}
\citet{Goldsmith2012} calculated that
\begin{equation}
    \label{eq:Rul_H2}
    R_{ul}({\rm H}_2) = 3.8\times10^{-10}\left(\frac{T}{100}\right)^{0.14} {\rm cm^3~s^{-1}}
\end{equation}
where $T$ is again the kinetic temperature. Using the same assume temperature and density as above, $R_{ul}({\rm H}_2) = 4.5\times10^{-10}~{\rm cm^3~s^{-1}}$, $\Sigma R_{ul,i}n_i = 4.5\times10^{-8}$~s\per, and $B = 2.9\times10^{-2}$. With these assumptions, Equation \ref{eq:crawford} becomes
\begin{equation}
    \label{eq:crawford_simp_H2}
    N_{\rm C^{+}}^{\rm H_{2}} = 1.0\times10^{17}\left[\frac{I_{\rm [CII]}}{\rm K}\right] \left[\frac{\Delta V}{\rm km~s^{-1}}\right] {\rm\ cm^{-2}}
\end{equation}
considering only collisions with the molecular gas.

From \NCpHt, we can estimate the effective \htwo\ column density (\NHtCII) based on the relative abundance of carbon to \htwo\ (${\rm C/H_2} = 7.5\times10^{-5}$; i.e., half the C/H ratio from \citealt{Gerin2015}) assuming all of the carbon is singly-ionized and all the hydrogen is molecular:
\begin{equation}
    \label{eq:N_H2_CII}
    N_{\rm H_{2}}^{\rm [CII]} = \frac{N_{\rm C^{+}}^{\rm H_{2}}}{{\rm C/H_2}}.
\end{equation}
\change{We show this effective \htwo\ column density profile based on the \CII\ in Figure \ref{fig:columndensity_H2} (teal).} %We compare these to the measured \htwo\ column density profiles from the IRAM data (see e.g., Section \ref{ssec:molgas_outflow}) in gray. In all pixels, \NHtCII~$>$~\NHtCO, meaning there is insufficient molecular gas (traced by CO) to explain the \CII\ as purely arising from the molecular phase. In the upper right corner of each panel, we report the ratio of \NHtCO$/$\NHtCII\ at the peaks of each profile. The ratios are from $\approx0.2-0.5$ (excluding Pixels 4 and 5 where there is no significant \CII\ signal), meaning that 20-50\% of the \CII\ could be excited by the molecular ISM alone (excluding any contribution from CO-dark molecular gas). 

\subsection{C$^+$ Mass Estimate}
\label{app:mass}
From the \Cp\ column density calculated from Equation \ref{eq:crawford_simp} (since most of the \CII\ is excited through collisions with atomic gas), we calculate the total \Cp\ mass (\MCp) in each upGREAT pointing in the outflow. In each LFA pixel, 
\begin{equation}
    \label{eq:MCp}
    M_{\rm C^+} = 2.2\left[\frac{A_{\rm pix}}{\rm arcsec^2}\right]\left[\frac{d}{\rm Mpc}\right]^2\left[\frac{\sum_{v}{N_{\rm C^+}}}{\rm 10^{18} cm^{-2}}\right] {\rm\ M_\odot}
\end{equation}
where $A_{\rm pix}$ is the area of each LFA pixel ($\approx156$\arcsec$^2$), $d$ is the distance to the galaxy in Mpc, and $\sum_{v}{N_{\rm C^+}}$ is the sum of the \Cp\ column density over all of the channels. We report \MCp\ in each LFA pixel in the pointing along the outflow in Table \ref{tab:outflowfits}. The total \Cp\ mass measured in these observations of the outflow is $4.2\times10^4$~\msun\ (excluding Pixels 4 and 5 where the \CII\ line is not detected). \MCp\ changes by a factor of $\sim2$ for a factor of 2 change in either the assumed CNM temperature ($T$) or density ($n_{\rm CNM}$).

\end{document}